\documentclass[final,3p,times]{elsarticle}

\usepackage{graphicx}
\usepackage{epstopdf}
\usepackage{epsfig}
\usepackage{amssymb,amsmath,stmaryrd,tabularx}
\usepackage{wrapfig}
\usepackage{bm}
\usepackage[T1]{fontenc}
\usepackage[center]{subfigure}
\usepackage[toc,page]{appendix}
\usepackage{float}
\usepackage{booktabs}
\usepackage{array}
\usepackage{blindtext}
\usepackage{amsthm}
\usepackage[dvipsnames]{xcolor}
\usepackage[normalem]{ulem}
\usepackage[percent]{overpic}
\usepackage{multirow}
\usepackage{hyperref}
\usepackage{amsfonts}
\usepackage{bbm}
\usepackage{ulem}

\renewcommand{\cite}{\citep}
\setcitestyle{authoryear}

\renewcommand{\d}{\partial}
\renewcommand{\L}{\mathcal L}
\newcommand{\e}{\epsilon}

\newcommand{\I}{{\mathbbm{i}}}

\let\vec\mathbf

\renewcommand{\[}{\begin{equation}}
\renewcommand{\]}{\end{equation}}

\newcommand{\C}{\mathcal C}
\newcommand{\J}{\mathcal J}

\newcommand{\G}{\mathcal G}

\newcommand{\avg}[1]{\left \langle #1 \right \rangle}

\begin{document}

\title{A phase field crystal theory of the kinematics of dislocation lines}

\definecolor{lightblue}{RGB}{51, 204, 255}
\definecolor{pinkish}{RGB}{255, 51, 204}
\definecolor{darkgreen}{RGB}{0, 128, 0}
\definecolor{darkblue}{RGB}{0, 0, 128}

\date{\today}
\author[OSLO]{Vidar Skogvoll}
\author[OSLO]{Luiza Angheluta}
\author[OSLO]{Audun Skaugen}
\author[TUD,DCMS]{Marco Salvalaglio}
\author[MIN]{Jorge Vi\~nals}

\address[OSLO]{PoreLab, Njord Centre, Department of Physics, University of Oslo, P. O. Box 1048, 0316 Oslo, Norway}
\address[TUD]{Institute of Scientific Computing, TU Dresden, 01062 Dresden, Germany}
\address[DCMS]{Dresden Center for Computational Materials Science (DCMS), TU Dresden, 01062 Dresden, Germany}
\address[MIN]{School of Physics and Astronomy, University of Minnesota, Minneapolis, MN 55455
}

\begin{abstract}
We introduce a dislocation density tensor and derive its kinematic evolution law from a phase field description of crystal deformations in three dimensions. 
The phase field crystal (PFC) model is used to define the lattice distortion, including topological singularities, and the associated configurational stresses. 
We derive an exact expression for the velocity of dislocation line determined by the phase field evolution, and show that dislocation motion in the PFC is driven by a Peach-Koehler force. 
As is well known from earlier PFC model studies, the configurational stress is not divergence free for a general field configuration. 
Therefore, we also present a method (PFCMEq) to constrain the diffusive dynamics to mechanical equilibrium by adding an independent and integrable distortion so that the total resulting stress is divergence free. 
In the PFCMEq model, the far-field stress agrees very well with the predictions from continuum elasticity, while the near-field stress around the dislocation core is regularized by the smooth nature of the phase-field. We apply this framework to study the rate of shrinkage of an dislocation loop seeded in its glide plane. 
\end{abstract}

\maketitle

\section{Introduction}

Plasticity in crystalline solids primarily refers to permanent deformations resulting from the nucleation, motion, and interaction of extended dislocations. 
Classical plasticity theories deal with the yielding of materials within continuum solid mechanics \cite{hillMathematicalTheoryPlasticity1998,wuContinuumMechanicsPlasticity2004}. 
Deviations from elastic response are described with additional variables (e.g., the plastic strain), which effectively describe the onset of plasticity (yielding criteria), as well as the mechanical properties of plastically deformed media (e.g., work hardening). 
A macroscopic description of the collective behavior of dislocation ensembles is thus achieved, usually assuming homogeneous media for large systems. 
In crystal plasticity, inhomogeneities and anisotropies are accounted for, with the theory having been implemented as a computationally efficient finite element model \cite{rotersOverviewConstitutiveLaws2010,pokharelPolycrystalPlasticityComparison2014}. 
These theories are largely phenomenological in nature, and rely on constitutive laws and material parameters to be determined by other methods, or extracted from experiments. 
They can be finely tuned, but sometimes fail in describing mesoscale effects \cite{rollettUnderstandingMaterialsMicrostructure2015}.
On the other hand, remarkable mesoscale descriptions have been developed by tracking single dislocations  \cite{kubinDislocationMicrostructuresPlastic1992,bulatovConnectingAtomisticMesoscale1998,sillsFundamentalsDislocationDynamics2016,koslowskiPhasefieldTheoryDislocation2002,rodneyPhaseFieldMethods2003}. 
These approaches typically evolve dislocation lines through Peach-Koehler type forces while incorporating their slip system, mobilities, and dislocation reactions phenomenologically. 
Stress fields are described within classical elasticity theory \cite{andersonTheoryDislocations2017}. 
Since linear elasticity predicts a singular elastic field at the dislocation core, theories featuring its regularization are usually exploited. 
Prominent examples are the non-singular theory obtained by spreading the Burgers vector isotropically about dislocation lines \cite{caiNonsingularContinuumTheory2006}, and the stress field regularization obtained within a strain gradient elasticity framework \cite{lazarNonsingularStressStrain2005}. 
Plastic behavior then emerges when considering systems with many dislocations and proper statistical sampling \cite{devincreDislocationMeanFree2008}. 
Still, the accuracy and predictive power of these approaches depend on how well dislocations are modeled as isolated objects. 
In this context, mesoscale theories that require a limited set of phenomenological inputs are instrumental in connecting macroscopic plastic behavior to microscopic features of crystalline materials.

The Phase Field Crystal (PFC) model is an alternative framework to describe the nonequilibrium evolution of defected materials at the mesoscale \cite{elderModelingElasticityCrystal2002,emmerichPhasefieldcrystalModelsCondensed2012,momeniMultiscaleFrameworkSimulationguided2018}. 
Within the phase field description, complex processes such as dislocation nucleation \cite{skogvollDislocationNucleationPhasefield2021}, dislocation dissociation and stacking fault formation \cite{mianroodiAtomisticallyDeterminedPhasefield2015}, creep \cite{berryAtomisticStudyDiffusionmediated2015}, fracture \cite{liuNanoscaleStudyNucleation2020}, and boundary driven grain motion \cite{provatasUsingPhasefieldCrystal2007,wuPhaseFieldCrystal2012,yamanakaPhaseFieldCrystal2017,salvalaglioDefectsGrainBoundaries2018} have been studied. 
The phase field allows a short scale regularization of defect core divergences inherent in classical elasticity, while allowing for the treatment of defect topology, grain boundary structures, and associated mobilities. 
For static studies, the only constitutive input required is the (defect free) equilibrium free energy, functional of the phase field, which has a minimizer that corresponds to a spatially periodic configuration. 
For time dependent problems, the phase field is generally assumed to obey a gradient flow that minimizes the free energy functional. 
When topological defects are present in the phase field configuration, their motion directly follows from the gradient flow, without any additional specification of slip systems, stacking fault energies, and line or boundary mobilities. 
The PFC model thus begins with the definition of a scalar order parameter (or phase field) $\psi(\mathbf{r},t)$, function of space and time, so that its equilibrium configuration corresponds to a perfectly periodic, undistorted, configuration.
A non-convex free energy functional $F[\psi]$ of the field and its gradients is chosen so that its minimizer has the same spatial symmetry as the lattice of interest \cite{elderAmplitudeExpansionBinary2010}. 
The requisite free energies have been derived by using the methods of density functional theory \cite{elderPhasefieldCrystalModeling2007,huangPhasefieldcrystalDynamicsBinary2010,archerDerivingPhaseField2019}, although our calculations below will rely on modified forms of the classical Brazovskii functional description of modulated phases \cite{brazovskiiPhaseTransitionIsotropic1975}, also known as the Swift-Hohenberg model in the convection literature \cite{swiftHydrodynamicFluctuationsConvective1977}.

Despite the model's successes to date, a clear connection with classical theory of dislocation motion in crystalline solids is lacking. 
At its most basic level, the phase field does not carry mass, and hence momentum. 
Therefore the only stresses (momentum current) that appear in the theory are the reversible contributions that arise from variations of the free energy with respect to distortions of the phase field \cite{skaugenDislocationDynamicsCrystal2018}. 
Neither momentum currents that arise in a material due to Galilean invariance, nor dissipative currents that would couple directly to the material distortion are present \cite{re:forster75}. 
Unlike classical theories of dislocation motion, the primary object of the model is the phase field, from which other quantities are derived. 
For an appropriate choice of the free energy functional, the phase field minimizer is a \lq\lq crystalline" phase in that translational symmetry is broken. 
As is conventionally the case, the minimizer admits an expansion in a reciprocal space basis set. 
This expansion is further restricted to include only those wave vector modes in reciprocal space that are critical at onset of the broken symmetry phase.
Configurational distortions of the phase field appear as slow (in space and time) modulations of the complex amplitudes of the expansion. 
A displacement vector is defined from the phase of the modulation.
Configurational topological defects are possible and appear as (combinations of) zeros of the complex amplitudes, points at which the phases of the modulation are singular. 
The corresponding defect current, however, is solely related to the phase field, and to the equation governing its temporal evolution. 
This is in contrast with more general dislocation density currents in solid mechanics which also include dissipative contributions. 
An attempt to bridge the PFC description and a field theory of dislocation {\em mechanics} has been given in Ref. \cite{acharyaFieldDislocationMechanics2020}, where an extended free energy is introduced, which includes a material elastic contribution and the coupling between the two.

Since the theory lacks a proper description of momentum conservation, it also cannot describe the relaxation of elastic excitations.
The first attempt at extending the PFC model to include elastic interactions considered a phenomenological second order temporal derivative in the equation of motion for the phase field \cite{stefanovicPhasefieldCrystalsElastic2006}, which allowed for fast relaxation of short-wavelength elastic disturbances. 
Later efforts have included coupling the PFC phase field to a velocity field \cite{ramosDynamicalTransitionsSliding2010}, or various methods of coarse graining it to develop a consistent hydrodynamical description \cite{tothNonlinearHydrodynamicTheory2013,heinonenConsistentHydrodynamicsPhase2016}. 
Such approaches are necessary for a proper description of processes where elastic interactions are important, such as crack propagation and defect dynamics. 
Other efforts have been made to develop efficient modeling approaches in which the time scale of elastic interactions is \textit{a priori} set to zero $\tau_E=0$, i.e. when mechanical equilibrium is obeyed at all times. 
The latter approach is justified when deformations are slow, including many of the applications mentioned such as creep and boundary driven grain motion. 
This approach was first introduced in Ref.~\cite{heinonenPhasefieldcrystalModelsMechanical2014} which involved relaxing elastic excitations separately and instantaneously within the amplitude equation formulation of the PFC model \cite{goldenfeldRenormalizationGroupApproach2005}. 
The same strategy was later developed for the PFC model in two dimensional isotropic 2D lattices by adding to the phase-field a correction at each time step that ensured instantaneous mechanical equilibrium \cite{skaugenSeparationElasticPlastic2018,salvalaglioCoarsegrainedPhasefieldCrystal2020}. 
In this paper, we present a generalization of this approach to anisotropic crystals in three dimensions (PFC-MEq). 
Since a distorted phase field configuration determines the corresponding configurational stresses \cite{skaugenDislocationDynamicsCrystal2018,skogvollStressOrderedSystems2021}, the method yields regularized stress profiles for dislocation lines in three dimensions down to the defect core. 
In the case of a point defect, it was shown in Ref.~\cite{salvalaglioCoarsegrainedPhasefieldCrystal2020} that the stress field at the core agrees with the predictions of the non-singular theory of Ref.~\cite{caiNonsingularContinuumTheory2006}, and with gradient elasticity models \cite{lazarNonsingularStressStrain2005,lazarNonsingularDislocationContinuum2017}, indicating that the results obtained here can serve as benchmarks for similar theories in three dimensions. 
The specific example of a dislocation loop in a bcc lattice is considered, and the far-field stresses given by the $\psi$ field are shown to coincide with predictions by continuum elasticity. 

The rest of the paper is structured as follows. 
In Sec. \ref{sec:DD_VEL}, we introduce the theoretical method used to define topological defects from a periodic $\psi$-field. 
This allows us to define a dislocation density tensor in terms of the phase field (Eq.~\eqref{eq:dislocation_density}), and obtain the dislocation line velocity (Eq.~\eqref{eq:3Ddislocationvelocity}).
These are key results, which are applied in several examples in Sec. \ref{sec:PFC}. 
First, we use the PFC model to numerically study the shrinkage of a dislocation loop in a bcc lattice. 
Then, we show analytically that Eq.~\eqref{eq:3Ddislocationvelocity} captures the motion of dislocations driven by a Peach-Koehler type force, and hence by a local stress. 
Finally, we introduce the PFC-MEq model, and compare the shrinkage of the dislocation loop under PFC and PFC-MEq dynamics. 
While the results are qualitatively similar for the case of a shear dislocation loop, the constraint of mechanical equilibrium causes the shrinkage to happen much faster. 
We finally confirm that the stress field derived from the $\psi$ field in the PFC-MEq model agrees with that which would follow from continuum elasticity theory, with the same singular dislocation density as source, and with no adjustable parameters.

\section{Kinematics of a dislocation line in three dimensions}
\label{sec:DD_VEL}

\begin{figure}
    \centering
    \includegraphics[]{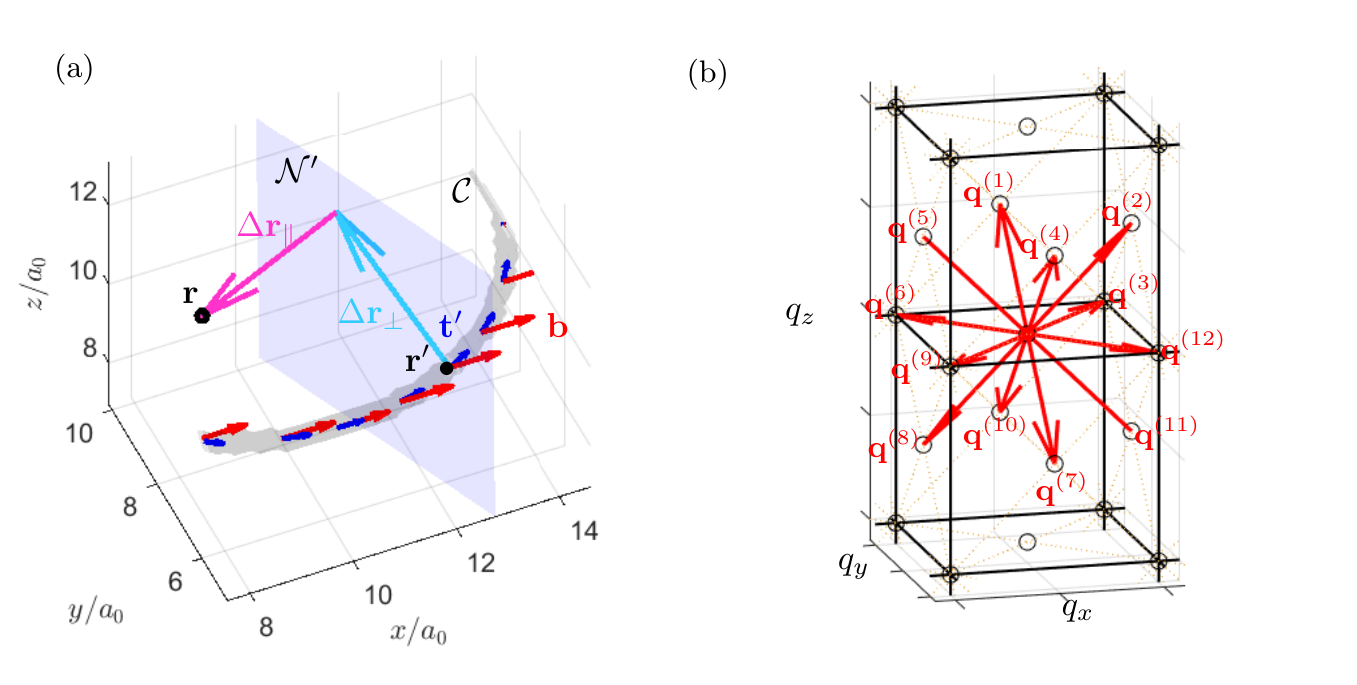}
    \caption{
    (a) A dislocation line $\mathcal C$ consisting of points $\vec r'$ characterized by the tangent vector $\vec t'$ and the Burgers vector $\vec b$ at that point.
    The difference $\vec r-\vec r'$ from a point $\vec r$ to a point $\vec r'$ on the line can be decomposed into a 2D in-plane vector $\Delta \vec r_\perp$, which is the projection of $\vec r-\vec r'$ onto the plane $\mathcal N'$ normal to $\vec t'$ and a distance $|\Delta \vec r_\parallel|$ from this plane.
            In this figure, $\Delta \vec r_\parallel \cdot \vec t' = -3.47 a_0$ and $|\Delta \vec r_\perp| = 3.42 a_0$.  
    (b) The $N=12$ primary reciprocal lattice vectors $\{\vec q^{(n)} \}_{n=1}^{12}$ of length $q_0$ of a bcc lattice (Eq. \eqref{eq:qn_bcc}).
    Higher modes (dots) correspond to higher harmonics $\{\vec p_n\}_{n>N}$ in the expansion of the equilibrium phase-field configuration $\psi^{eq}$, Eq. \eqref{eq:equilibrium_psi_field}. 
    }
    \label{fig:coordinate_decomposition}
\end{figure}

Dislocations in 3D crystals are line defects, where each point $\vec r'$ on the line $\mathcal C$ is characterized by the tangent vector $\vec t'$ at that point and a Burgers vector $\vec b$, see Figure  \ref{fig:coordinate_decomposition}(a).
By introducing a local Cartesian plane $\mathcal N'$ normal to $\vec t'$, the distance of an arbitrary point $\vec r$ to a point $\vec r'$ on $\C$ can be decomposed into an in-plane vector $\Delta \vec r_\perp \perp \vec t'$ and a vector $\Delta \vec r_\parallel \parallel \vec t'$, i.e. $\vec r -\vec r' = \Delta \vec r_\perp  + \Delta \vec r_\parallel$.
A deformed state can be described by a displacement field $\vec u$ and, in the presence of a dislocation, $\vec u$ is discontinuous across a surface (branch cut) spanned by the dislocation, given by 
\[
\oint_{\Gamma'} d\vec u = \vec u^+ - \vec u^{-} = -\vec b,
\label{eq:Burgers_vector_definition}
\]
where $\vec u^+$ and $\vec u^-$ are the values of the displacement field at each side of the branch cut, respectively. 
We use the negative sign convention relating the contour integral with the Burgers vector. 
Here, $\Gamma'$ is a small circuit enclosing the dislocation line in the $\mathcal N'$-plane, directed according to the right-hand rule with respect to $\vec t'$. 
The dislocation density tensor associated with the line is \cite{lazarGradientFieldTheories2014}
\begin{equation}\label{eq:alpha_ik_definition_in_terms_of_ti_bk}
\alpha = \boldsymbol \delta^{(2)}(\C)\otimes \vec b  = \left (\int_{\C} d\vec l'\delta^{(3)} (\vec r-\vec r' )\right )  \otimes \vec b
\qquad \left (
\alpha_{ij} (\vec r) =  b_j \delta_i^{(2)} (\C) = b_j \int_{\C} dl_i' \delta^{(3)} (\vec r-\vec r' )
\right ),
\end{equation}
where $b_{j}$ is the $j$ component of the Burgers vector of the line, and $dl'_i =  t'_i dl'$ is the line element in the direction of the line.  $\delta^{(2)}_i(\C)$ is a short-hand notation for the delta function, with dimension of inverse area, locating the position of the dislocation line for each component $i$ of the dislocation density tensor. 
It is defined by the line integral over the dislocation line of the full delta function (which scales as inverse volume). 
The dislocation density tensor is defined so that $\int_{\mathcal N'} d^2 r_\perp \alpha_{ij} t'_i = b_j$, where we are using the Einstein summation convention over repeated indices. 

In the PFC models, a crystal state is represented by a periodic phase field $\psi(\vec r)$ of a given crystal symmetry. 
A reference crystalline lattice, 
\[\psi^{eq}(\vec r) = \bar \psi + \sum_{n=1}^N\eta_{n} e^{\mathbbm i \vec q^{(n)} \cdot \vec r} + \sum_{n>N} \eta_{n}e^{\mathbbm i \vec p^{(n)} \cdot \vec r},
\label{eq:equilibrium_psi_field}
\]
is defined by a set of $N$ primary (smallest) reciprocal lattice vectors $\{\vec q^{(n)} \}_{n=1}^N$ of length $q_0$, and higher harmonics $\{\vec p_n\}_{n>N}$, also on the reciprocal lattice but with $|\vec p_n|>q_0$ (see, e.g., $\{\vec q^{(n)}\}$ with $|\vec q^{(n)}|=q_0$ for a bcc lattice in Fig.~\ref{fig:coordinate_decomposition}(b)). 
The lattice constant of the crystal is then given by $a_0 \sim 2\pi/q_0$.
This represents a perfect crystal configuration in the absence of defects and distortion, where the average value $\bar \psi$ and the amplitudes $\eta_n$ are constants.
In the phase-field crystal theory presented in Refs. \cite{elderModelingElasticityCrystal2002,elderModelingElasticPlastic2004}, near the solid-liquid transition point, only the terms from the primary reciprocal lattice vectors contribute to $\psi^{eq}$, while in general for more sharply peaked density profiles, there are also contributions from the higher order harmonics $\{ \vec p_n\}_{n>N}$. 
For a distorted crystal lattice, the mode amplitudes $\eta_n$ become complex scalar fields, henceforth named \textit{complex amplitudes} $\eta_n(\vec r)$, such that
\[\psi (\vec r) \approx \bar \psi (\vec r) + \sum_{n=1}^N \eta_n(\vec r) e^{\mathbbm i \vec q^{(n)} \cdot \vec r} +  \sum_{n>N} \eta_{n} (\vec r) e^{\mathbbm i \vec p^{(n)} \cdot \vec r}.\]
 In this section, we provide an accurate description of dislocation lines as topological defects in the phase of the complex amplitudes $\eta_n(\vec r)$. 
 We generalize the method of tracking topological defects as zeros of a complex order parameter as introduced in Refs. \cite{halperinStatisticalMechanicsTopological1981} and \cite{mazenkoVortexVelocitiesSymmetric1997}, and apply it to accurately derive the kinematics of dislocation lines. 

\begin{figure}[]
    \centering
    \includegraphics[]{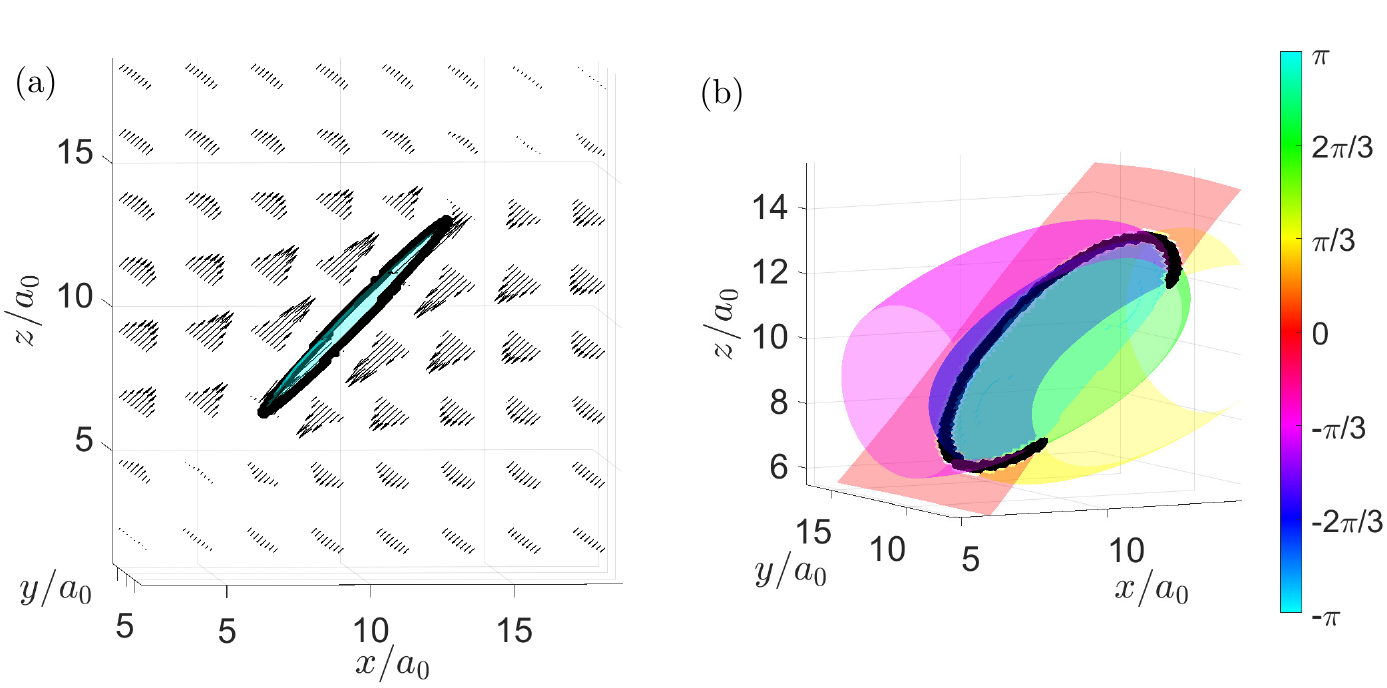}
    \caption{ 
    (a) Example of a discontinuous displacement field $\vec u$ in the presence of a dislocation loop (black line) with Burgers vector $\vec b$ (Eq. \eqref{eq:Burgers_vector_definition}).
    (b) Isosurfaces of one of the phases $\theta_n$ possessing the same discontinuity as the displacement field (Eq. \eqref{eq:thetan}). }
    \label{fig:amplitude_discontinuity}
\end{figure}

Given a phase field configuration $\psi(\vec r)$, the complex amplitudes can be found by a demodulation as described in \ref{appendix:amplitude_demodulation}.
Decomposing each amplitude $\eta_n(\vec r) = \rho_n (\vec r) e^{i\theta_n (\vec r)}$, into its modulus $\rho_n(\vec r)$ and phase $\theta_n(\vec r)$, we have that for a perfect lattice, $\theta^{(0)}_n =0$ and $\rho_n$ is constant.
Displacing a lattice plane by a slowly varying $\vec u$ transforms the phase as $\theta_n \rightarrow \theta_n^{(0)} - \vec q^{(n)} \cdot \vec u$. 
Thus, the phase provides a direct measure of the displacement field $\vec u(\vec r)$ relative to the reference lattice, i.e.,
\begin{equation}
\label{eq:thetan}
\theta_n (\vec r) = -\vec q^{(n)} \cdot \vec u(\vec r) 
\qquad
\left (\theta_n (\vec r) = -q_{i}^{(n)} u_i(\vec r)
\right ),
\end{equation}
where $q_{i}^{(n)}$ denotes the $i$-th Cartesian coordinate of $\vec q^{(n)}$. 
It is possible to invert Eq.~(\ref{eq:thetan}), and solve for the displacement field $\vec u$ as function of the phases $\theta_n$ and reciprocal vectors. 
We use the following identity which is valid for lattices with cubic symmetry, where all primary reciprocal lattice vectors have the same length $q_0$ (see  \ref{appendix:inversion_identity})
\[
\sum_{n=1}^N \vec q^{(n)} \otimes \vec q^{(n)} = \frac{Nq_0^2}{3} \mathbbm 1
\qquad
\left (
\sum_{n=1}^N q_{i}^{(n)} q_{j}^{(n)} = \frac{Nq_0^2}{3}  \delta_{ij}
\right ),
\label{eq:Qij_identity}
\]
so that the displacement $\vec u$ is given by 
\[
\vec u(\vec  r) = -\frac{3}{N q_0^2} \sum_{n=1}^N \vec q^{(n)}\theta_n (\vec r).\label{eq:u}
\]
Eq. \eqref{eq:u} shows that a dislocation line, which introduces a discontinuity in the displacement field, leads to a discontinuity in the phases $\theta_n(\vec r)$. 
This is the first key insight, which we illustrate in Fig.~\ref{fig:amplitude_discontinuity}.
By using Eq.~(\ref{eq:thetan}) and the fact that the Burgers vector $\vec b$ is constant along the dislocation line, we relate the Burgers vector to the phase $\theta_n$ as
\begin{equation}\label{eq:s_n}
\oint_{\Gamma'} d \theta_n  = -\oint_{\Gamma'} \vec q^{(n)} \cdot d\vec u
= \vec q^{(n)} \cdot \vec b \equiv 2\pi s_{n},
\end{equation}
where $s_{n} $ is the (integer) winding number of the phase $\theta_n$ around the dislocation line. 
That $s_n$ is an integer follows from the fact that while $\theta_n(\vec r)$ may have a discontinuity across the branch cut, the complex amplitude $\eta_n(\vec r)$ is well-defined and continuous everywhere. 
Therefore the circulation of the phase must be an integer multiple of $2\pi$. 
By the same reasoning, for an amplitude for which $\vec q^{(n)} \cdot \vec b \neq 0$, at the dislocation line, the phase $\theta_n(\vec r)$ is undefined (singular), so the modulus $\rho_n(\vec r)$ must go to zero for $\eta_n(\vec r)$ to remain continuous. 
This is the second key insight, which allows us to identify the location of the dislocation line with the zeros of the complex amplitudes $\eta_n(\vec r)$.  

The complex amplitude $\eta_n(\vec r)$ is isomorphic to a $2$-component vector field $\vec \Psi(\vec r) \equiv (\Psi_1(\vec r),\Psi_2(\vec r)) = (\Re (\eta_n(\vec r)),\Im(\eta_n(\vec r)))$. 
The study of how to track zeros of any dimensional vector field in any dimensions was introduced in Ref. \cite{halperinStatisticalMechanicsTopological1981}. 
The orientation field $\vec \Psi(\vec r)/|\vec \Psi(\vec r)|$ is continuous wherever $|\vec \Psi(\vec r)|\neq 0$ and supports 1D topological defects in 3 dimensions which are located precisely where $|\vec \Psi(\vec r)|=0$. 
The topological line density $\rho_i$ of the line $\mathcal C$, which satisfies $\int d^2r_\perp \rho_i = s_n t_i'$, is given by  
\[
\boldsymbol \rho = s_n \boldsymbol \delta^{(2)}(\C) 
\qquad 
\left (
\rho_i = s_n \delta^{(2)}_i(\C)
\right ).
\]
Like $\delta^{(2)}_i(\mathcal C) $, the dimension of $\rho_i$ is that of a two-dimensional vector density. 
This topological charge density is expressed explicitly in terms of the real-valued positions $\mathcal C = \{\vec r'\}$ of the topological defect line.
Since these positions coincide with the zero-line of the vector field $\vec \Psi(\vec r)$, it is possible to relate the expression to the delta-function locating the zeros of $\vec \Psi(\vec r)$, through the transformation law $ s_n \delta^{(2)}(\mathcal C) = D_i(\vec r) \delta^{(2)} (\vec \Psi(\vec r))$, with the determinant vector field $D_i(\vec r) = \epsilon_{ijk} (\partial_j \Psi_1(\vec r)) (\partial_k \Psi_2(\vec r))$. 
Comparing this to Eq. \eqref{eq:alpha_ik_definition_in_terms_of_ti_bk}, using Eq. \eqref{eq:s_n} and re-expressing $D^{(n)}_i(\vec r)$ (with the added superscript $n$) in terms of the complex amplitude $\eta_n(\vec r)$, we end up with the central equation for tracking the evolution of the dislocation density
\[
\frac{1}{2\pi}  \vec q^{(n)} \cdot \alpha(\vec r) = \delta^{(2)}(\eta_n(\vec r)) \vec D^{(n)} (\vec r)
\qquad 
\left (
\frac{1}{2\pi} q_{k}^{(n)} \alpha_{ik}(\vec r) 
=
\delta^{(2)}(\eta_n(\vec r)) D^{(n)}_i(\vec r)
\right ),
\label{eq:alpha_tensor_amplitude_relationship}
\]
where $\delta^{(2)}(\eta_n) = \delta(\Re(\eta_n)) \delta(\Im(\eta_n))$ and 
\[
\vec D^{(n)}(\vec r) = \nabla \Re(\eta_n(\vec r)) \times \nabla \Im(\eta_n(\vec r))
\qquad \left (
 D^{(n)}_{i}(\vec r) = \epsilon_{ijk} (\partial_j \Re(\eta_n(\vec r))) (\partial_k\Im(\eta_n(\vec r)))
\right ).
\label{eq:determinant_vector}
\]
In the following, for ease of notation, we suppress the explicit positional dependence of $\alpha_{ij},D_i^{(n)}$ and $\eta_n$. 
The dislocation line is located at $\eta_n = 0$, which is the intersection of the surfaces $\Re(\eta_n)=0$ and $\Im(\eta_n)=0$.
As we see from its definition, $\vec D^{(n)}$ is perpendicular to both these surfaces and is thus directed along the tangent to the line. 
We can reconstruct the dislocation density tensor from an appropriate summation over the modes with singular phases, namely by multiplying Eq. \eqref{eq:alpha_tensor_amplitude_relationship} by $q_j^{(n)}$, summing over the reciprocal modes and using Eq.~(\ref{eq:Qij_identity}) to arrive at
\begin{equation}
\alpha = \frac{6 \pi}{N q_0^2} \sum_{n=1}^N \delta^{(2)}(\eta_n) \vec D^{(n)} \otimes \vec q^{(n)} 
\qquad
\left (
\alpha_{ij} = \frac{6 \pi}{N q_0^2} \sum_{n=1}^N\delta^{(2)}(\eta_n)   D^{(n)}_i  q_{j}^{(n)}
\right ).
\label{eq:dislocation_density}
\end{equation}

Having a closed form of the dislocation density in terms of the complex amplitudes $\eta_n$, we now turn to deriving a closed form expression for its kinematic in terms of the time evolution of $\eta_n$. 
Taking the time derivative of Eq. \eqref{eq:alpha_ik_definition_in_terms_of_ti_bk}, we show in \ref{appendix:derivative_of_dislocation_density_tensor_delta_function_form} that for a dislocation density tensor described by a single loop or string, we have $\partial_t \alpha_{ij} = -\epsilon_{ikl} \partial_k  \J_{lj}^{(\alpha)}$, where \[
\mathcal J^{(\alpha)} = \alpha \times \vec V
\qquad \left (
\J_{lj}^{(\alpha)} = \epsilon_{lmn} \alpha_{mj} V_n \right ),\]
and $\vec V$ is a vector field defined on the string by the velocity of the line segment perpendicular to the tangent vector.
Taking the time derivative of Eq. \eqref{eq:dislocation_density}, we show in \ref{appendix:derivative_of_dislocation_density_tensor_amplitude_form} that we get $\partial_t \alpha_{ij} =- \epsilon_{ikl} \partial_k \J_{lj}$, where 
\[
\J = \frac{6 \pi}{N q_0^2} \sum_{n=1}^N\delta^{(2)}(\eta_n) \vec J^{(n)} \otimes \vec q^{(n)} 
\qquad 
\left (
\J_{lj} = \frac{6 \pi}{N q_0^2} \sum_{n=1}^N  \delta^{(2)}(\eta_n) J_{l}^{(n)} q_{j}^{(n)}\right ),
\label{eq:dislocation_current_amplitudes}
\]
and $J_{l}^{(n)} = (\partial_l \Re(\eta_n))\partial_t\Im(\eta_n) - (\partial_l \Im(\eta_n))\partial_t\Re(\eta_n)  = \Im (\d_t \eta_n \d_l \eta_n^*)$.
Note that $\J_{lj}$ depends on $ \d_t \eta_n$, and hence on the law governing the temporal evolution of the phase field.
$\mathcal J^{(\alpha)}_{lj}$ is the well-known expression in terms of the dislocation velocity and $\mathcal J_{lj}$ is what we predict from the evolution of the phase field crystal density $\psi$. 
Under the assumption that both currents are equal, we show in the following that we are able to determine the dislocation velocity directly from the evolution of the phase field $\psi$ at the dislocation core.
We have checked numerically that the dislocation velocity predicted with this assumption is in excellent agreement with the one computed by tracking the position of the dislocation line at successive time steps.

By contracting Eq. \eqref{eq:alpha_tensor_amplitude_relationship} with $D_i^{(n)}$, we can express the delta-function in terms of the dislocation density tensor $\delta^{(2)}(\eta_n) = \alpha_{ik} D_i^{(n)}q_k^{(n)}/ (2\pi |\vec D^{(n)}|^2) $, which we can insert into Eq. \eqref{eq:dislocation_current_amplitudes}. 
Then, by equating $\mathcal J_{lj}$ and $\mathcal J^{(\alpha)}_{lj}$ at a point $\vec r'$ on the dislocation line, where $\alpha_{ik} = t'_i b_k \delta^{(2)}(\Delta \vec r_\perp)$, we get after contracting with $\vec b$ and integrating the delta-functions in $\mathcal N'$ (details in \ref{appendix:details_of_velocity_calculation}) 
\[
\frac{12 \pi^2}{N q_0^2 |\vec b|^2} \sum_{n=1}^N s_n^2 \left (\frac{\vec t' \cdot \vec D^{(n)}}{|\vec D^{(n)}|^2} \right ) \vec J^{(n)}  
=
\vec  t' \times \vec v'
\qquad
\left (
\frac{12 \pi^2}{N q_0^2 |\vec b|^2} \sum_{n=1}^N s_n^2 \frac{t_i' D_i^{(n)}}{|\vec D^{(n)}|^2} J_{l}^{(n)}  
=
\epsilon_{lmn} t_m'  v_n'
\right ),
\]
where $\vec v'$ is the velocity of the dislocation node at $\vec r'$. 
Since $\vec t' \perp \vec v'$, we can easily invert this relation to find $\vec v'$, and using that $\vec D^{(n)} \parallel \vec t'$ gives
\[
\vec v'
=
\frac{12\pi^2}{Nq_0^2 |\vec b|^2} \sum_{n=1}^N s_n^2    \frac{\vec J^{(n)} \times \vec  D^{(n)}}{|\vec D^{(n)}|^2}
\qquad 
\left (
v_s' = \frac{12\pi^2}{Nq_0^2 |\vec b|^2} \sum_{n=1}^N s_n^2    \frac{\epsilon_{slr} J_{l}^{(n)} D_r^{(n)}}{|\vec D^{(n)}|^2}
\right ).
\label{eq:3Ddislocationvelocity}
\]

Eqs.~(\ref{eq:dislocation_density}) and (\ref{eq:3Ddislocationvelocity}) are the key results of this paper. Eq.~(\ref{eq:dislocation_density}) defines the dislocation density tensor from the demodulated amplitudes $\eta_n$ of the phase field, while Eq.~\eqref{eq:3Ddislocationvelocity} gives an explicit expression for the dislocation line velocity. Both equations bridge the continuum description of the dislocation density and velocity with the microscopic scale of the phase field.

\section{Dislocation motion in a bcc lattice}
\label{sec:PFC}

\begin{figure}
    \centering
    \includegraphics[]{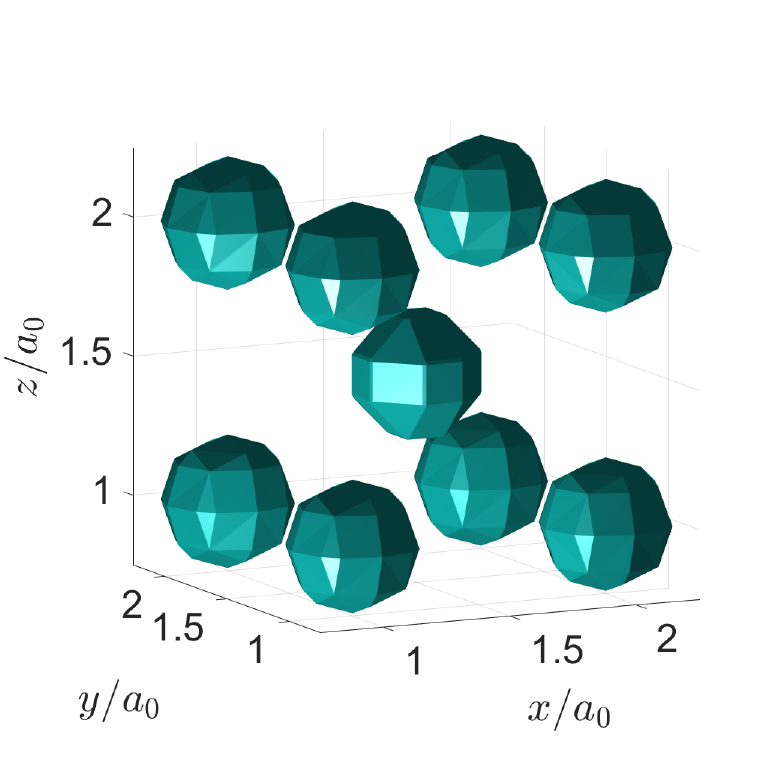}
    \caption{A bcc unit cell in the PFC model shown by isosurfaces of constant $\psi$. Using the model parameters given in ~\ref{appendix:numerical_methods}, $\psi(\vec r)$ varies between peaks of $\psi=0.7447$ and troughs of $\psi=-0.6148$ with the isosurfaces drawn at $\psi=0.0694$.}
    \label{fig:PFC_one_bcc_unit_cell}
\end{figure}

We apply here the framework developed in Sec.~\ref{sec:DD_VEL} to a phase field crystal model of dislocation motion in a bcc lattice \cite{elderModelingElasticityCrystal2002,elderModelingElasticPlastic2004,emmerichPhasefieldcrystalModelsCondensed2012}. 
The free energy $F_\psi$ is a functional of the phase field $\psi$ over the domain $\Omega$, given by
\begin{equation}
F_\psi=\int_{\Omega} \left[\frac{\Delta B_0}{2}\psi^2+\frac{B^x_0}{2} \psi\mathcal L^2\psi 
-\frac{T}{3}\psi^3+\frac{V}{4}\psi^4 \right]d\mathbf{r},
\label{eq:F_PFC}
\end{equation}
where $\mathcal L = q_0^2+\nabla^2$, and $\Delta B_0$, $B_0^x$, $V$, and $T$ are constant parameters  \cite{elderPhasefieldCrystalModeling2007}.
The dissipative relaxation of $\psi$ reads as
\[
\frac{\partial \psi}{\partial t}
=
\Gamma \nabla^2 \frac{\delta F_\psi}{\delta \psi}.
\label{eq:PFC_diffusive_dynamics}
\]
with constant mobility $\Gamma$. 
We will refer to Eq.~(\ref{eq:PFC_diffusive_dynamics}) as the "classical" PFC dynamics. 
As a characteristic unit of time given these model parameters, we use $\tau = (\Gamma B_0^x q_0^6)^{-1}$. 
For appropriate parameter values, the ground state of this energy is a bcc lattice which is well described in the one mode approximation 
\[
\psi(\vec r)= \psi_0 + \sum_{n=1}^{12} \eta_0  e^{\I\vec{q}^{(n)} \cdot \mathbf{r}},
\label{eq:psi_approx}
\]
where $\psi_0$ is the average density, $\eta_0$ is the equilibrium amplitude found by minimizing the free energy (Eq. \eqref{eq:F_PFC}) with this ansatz for $\psi(\vec r)$, and $\{\vec q_n\}$ are the $N=12$ smallest reciprocal lattice vectors
\[
\begin{array}{ll}
\vec q^{(1)} = q_0(0,1,1)/\sqrt 2, &  \vec q^{(4)} = q_0(0,-1,1)/\sqrt 2,\\
\vec q^{(2)} = q_0(1,0,1)/\sqrt 2, & \vec q^{(5)} = q_0(-1,0,1)/\sqrt 2, \\
\vec q^{(3)} = q_0(1,1,0)/\sqrt 2, & \vec q^{(6)} = q_0(-1,1,0)/\sqrt 2,\\
\end{array}
\label{eq:qn_bcc}
\]
with $\vec q^{(n)} = -\vec q^{(n-6)}$ for $n=7,..,12$, see Fig.~\ref{fig:coordinate_decomposition}(b).
Figure \ref{fig:PFC_one_bcc_unit_cell} shows one bcc unit cell of a phase-field initialized in the one-mode approximation.
Given the equilibrium configuration, the lattice constant $a_0$ will be used as the characteristic unit of length and the shear modulus $\mu$ calculated from the phase-field will serve as the characteristic unit of stress \cite{skogvollStressOrderedSystems2021}. 
As we see, the functional form of the free energy determines the base vectors $\vec q^{(n)}$, and no further assumptions about slip systems or constitutive laws for dislocation velocity (or plastic strain rates) need to be introduced. 

The model parameters ($\Delta B_0, B_0^x, T,V$, and $\Gamma$) and variables ($F_\psi,\psi,\vec r$, and $t$) can be rescaled to a dimensionless form in which $B_0^x=V=q_0=\Gamma=1$, thus leaving only three tunable model parameters: the quenching depth $\Delta B_0$, $T$ and the average density $\psi_0$ (due to the conserved nature of Eq. \eqref{eq:PFC_diffusive_dynamics}). 
All simulations are performed in these dimensionless units as described in Sec. \ref{appendix:numerical_methods_phase_field_evolution}.

\subsection{Numerical analysis: shrinkage of a dislocation loop}
\label{sec:numerical_loop_shrinkage}

In order to have a lattice containing one dislocation loop as the initial condition, we consider first the demodulation of the $\psi$ field in the one mode approximation. 
A dislocation loop is introduced into the perfect lattice by multiplying the equilibrium amplitudes by complex phases $\eta_0\rightarrow \eta_n(\vec r)$ with the appropriate charges $s_n$ (see \ref{appendix:initial_condition_explanation}) and then reconstructing the phase field $\psi$ through Eq.~\eqref{eq:psi_approx}. 
We then integrate Eq.~\eqref{eq:PFC_diffusive_dynamics} forward in time as detailed in ~\ref{appendix:numerical_methods_phase_field_evolution}. 
A fast relaxation follows from the initial configuration with the loop. 
This relaxation leads to the regularization of the singularity at the dislocation line ($\eta_n \rightarrow 0$ for $s_n \neq 0$) as achieved in PFC approaches \cite{skaugenDislocationDynamicsCrystal2018,salvalaglioClosingGapAtomicscale2019,salvalaglioCoarsegrainedPhasefieldCrystal2020}. 
From then onward, $\psi$ evolves in time leading to the motion of the dislocation line which may be analyzed by the methods outlined in Sec.~\ref{sec:DD_VEL}, using the amplitudes $\{\eta_n\}$ extracted from $\psi$ extracted as detailed in \ref{appendix:amplitude_demodulation}.

\begin{figure*}[htp]
    \centering
    \includegraphics[]{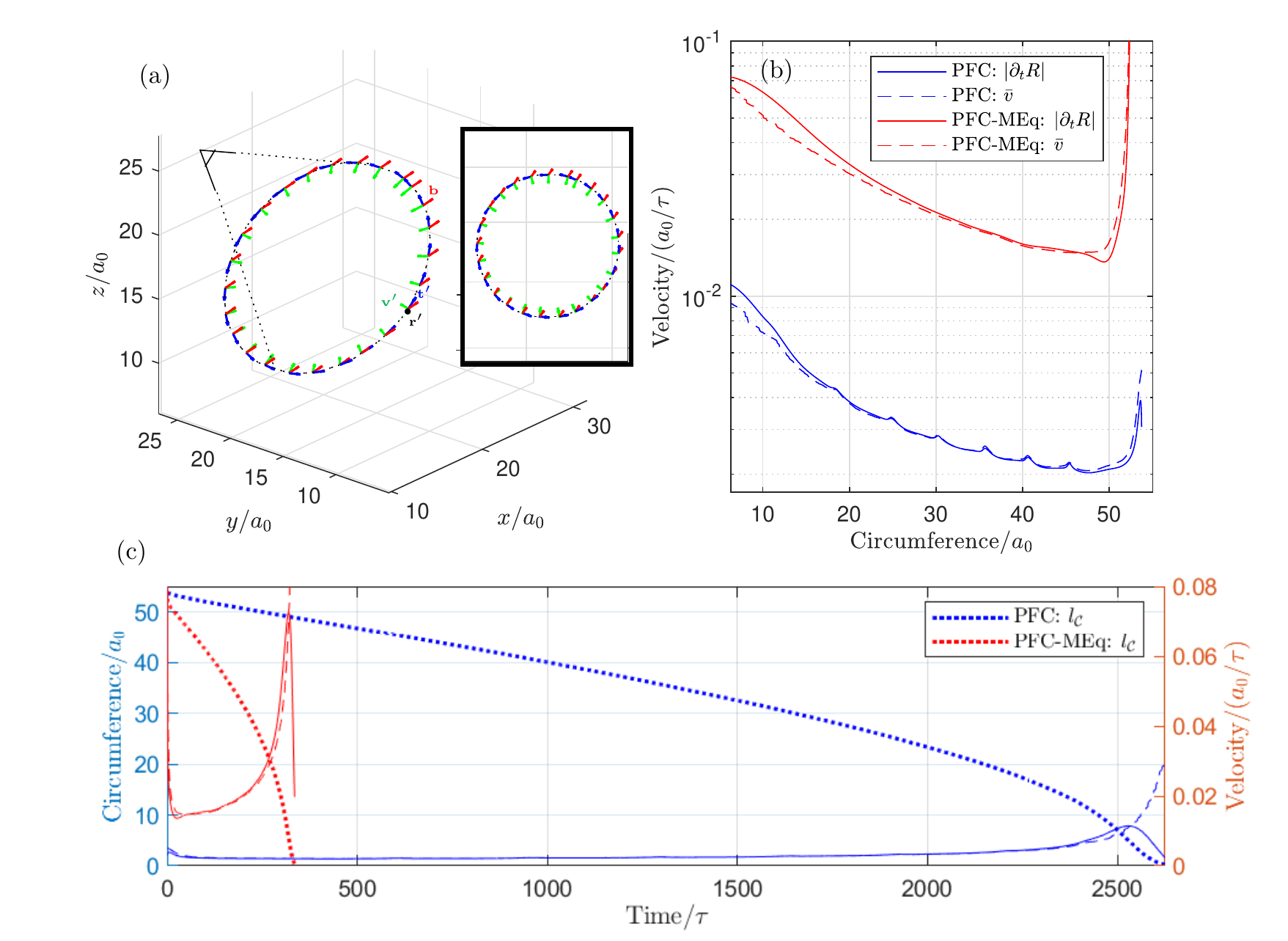}
    \caption{(a) The initial dislocation loop in a $35\times35\times35$ bcc PFC lattice with periodic boundary conditions. 
    The dislocation loop is on the slip system given by plane normal $[-1,0,1]$ and slip direction (Burgers vector) $\frac{a_0}{2} [1,-1,1]$. 
    Inset: the dislocation viewed from the indicated angle.
    (b) Comparison of average point velocity $\bar v$ (Eq. \eqref{eq:avg_point_velocity}) to the average loop radius shrinkage velocity $|\partial_t R|$ (Eq. \eqref{eq:avg_loop_radius_shrinkage}) as functions of the loop circumference.
    PFC and PFC-MEq refer, respectively, to the classical PFC model, and the PFC model constrained to mechanical equilibrium as introduced in Sec. \ref{sec:PFC_mechanical_equilibrium}.
    (c) The circumference $l_{\mathcal C}$ of the dislocation loop. 
    Superimposed on the right y-axis are the velocities of panel (b) as functions of time. 
    }
    \label{fig:loop_shrinkage}
\end{figure*}

Numerically, we approximate the delta function in Eq.~\eqref{eq:dislocation_density} as a sharply peaked 2D Gaussian distribution, i.e., $\delta^{(2)}(\eta_n) \simeq \exp(-\frac{| \eta_n|^2}{2\omega^2})/(2\pi \omega^2) $ with a standard deviation of $\omega = \eta_0/10$. 
Near the dislocation line, the dislocation density $\alpha_{ij}$ thus takes the form of a sharply peaked function, which can be treated numerically.
The decomposition of $\alpha_{ij}$ into its outer product factors $t_i'$ and a Burgers vector density $B_j = b_j \delta^{(2)}(\Delta \vec r_\perp)$ is done by singular value decomposition (see Sec.~\ref{appendix:SVD_of_alpha_in_tiBj}), and the Burgers vector of the point is extracted by performing a local surface integral in $\mathcal N'$. 
We prepare a $35\times35\times35$ unit cell 3D PFC lattice on periodic boundary conditions with a resolution of $\Delta x = \Delta y = \Delta z = a_0/7$. 
A dislocation loop is introduced as the initial condition in the slip system given by a plane normal $[-1,0,1]$ with slip direction (Burgers vector) $\frac{a_0}{2} [1,-1,1]$. 
Figure \ref{fig:loop_shrinkage}(a) shows the initial dislocation density decomposed as described, where we also have calculated the velocity $\vec v'$ at each point given by Eq~\eqref{eq:3Ddislocationvelocity}.

In order to obtain the velocity of the dislocation loop segments, we identify $M$ nodes on the loop and evaluate Eq.~(\ref{eq:3Ddislocationvelocity}) by using numerical differentiation of the $\psi$ field to calculate the amplitude currents $J_l^{(n)}$. 
To serve as a benchmark, we also calculate the circumference $l_{\C}$ of the dislocation loop $\C$ at each time (further details in \ref{appendix:numerical_methods_dislocation_perimeter}), so that we compare the rate of shrinkage $|\d_t R|$ 
\[
|\d_t R| = \frac{1}{2\pi} |\d_t l_{\C}|,
\label{eq:avg_loop_radius_shrinkage}
\]
(solid blue line in Fig.~\ref{fig:loop_shrinkage}(b)) to the average velocity of the $M$ dislocation nodes
\[
\bar v = \frac{1}{M} \sum_{m=1}^M |\vec v^{(m)}|, 
\label{eq:avg_point_velocity}
\]
(dashed blue line in Fig.~\ref{fig:loop_shrinkage}(b)) where $\vec v^{(m)}$ is the velocity of the dislocation line at node $m$, calculated by the velocity formula Eq.~(\ref{eq:3Ddislocationvelocity}). 
$|\d_t R| $ and $ \bar v $ should agree in the case of the shrinking of a perfectly circular loop and the figure shows excellent agreement between the two. 
Interestingly, we observe that both are sensitive to the Peierls like barriers during their motion, as shown by the oscillations in Fig.~\ref{fig:loop_shrinkage}(b)). 
The maxima are separated by $2\pi a_0$, confirming that the oscillation is related to the motion of a loop segment over one lattice spacing $a_0$ \cite{re:boyer02}. 
This observation confirms that even though Eqs.~\eqref{eq:dislocation_density} and \eqref{eq:3Ddislocationvelocity} are continuum level descriptions of the system, they still exhibit behavior related to the underlying lattice configuration.
The initial fast drop in velocity is due to the fast relaxation of the initial condition.  
The evolution of the variables under the dynamics of Eq. \eqref{eq:PFC_diffusive_dynamics} are shown together with the evolution given by the  PFC-MEq model which will be introduced in Sec. \ref{sec:PFC_mechanical_equilibrium}.

\subsection{Theoretical analysis: Peach Koehler law}
\label{sec:theoretical_PK_force}

In this section, we show that the general expression Eq.~\eqref{eq:3Ddislocationvelocity} of the defect velocity agrees with the dissipative motion of a dislocation as given by the classical Peach-Koehler force~\cite{re:pismen99,kosevichCrystalDislocationsTheory1979}. 
To calculate an analytical expression for the amplitude currents $J_l$, we employ the amplitude formulation of the PFC model, which directly expresses the free energy and dynamical equations in terms of the complex amplitudes $\eta_n$ \cite{goldenfeldRenormalizationGroupApproach2005,athreyaRenormalizationgroupTheoryPhasefield2006,salvalaglioCoarseGrainedModeling2022}. 
For our lattice symmetry, real valuedness of $\psi$ requires that $\eta_{n+6}=\eta_n^*$, and the dynamical equations need only consider the amplitudes $\{\eta_{n}\}_{n=1}^6$. 
By substituting Eq.~\eqref{eq:psi_approx} in $F_\psi$ and integrating over the unit cell, under the assumption of slowly-varying amplitudes, one obtains the following free energy as a function of the complex amplitudes,
\[
F_\eta=\int_{\Omega} 
\bigg[\frac{\Delta B_0}{2}\Phi+\frac{3V}{4}\Phi^2 +\sum_{n=1}^6
\left ( B_0^x |\mathcal{G}_n \eta_n|^2-\frac{3V}{2}|\eta_n|^4 \right )
+f^{\rm s}(\{\eta_n\},\{\eta^*_n\}) \bigg]  d \mathbf{r}, 
\label{eq:energyamplitude}
\]
where $\G_n = \nabla^2+2\I\mathbf{q}_n \cdot \nabla$ and $\Phi = 2\sum_{n=1}^6 |\eta_n|^2$. 
$f^{\rm s}(\{\eta_n\},\{\eta^*_n\})$ is a polynomial in $\eta_n$ and $\eta^*_n$ that depends in general on the specific crystalline symmetry under consideration \cite{goldenfeldRenormalizationGroupApproach2005,elderAmplitudeExpansionBinary2010,salvalaglioCoarseGrainedModeling2022} (here bcc, see  \ref{appendix:amplitude_decoupling} for its expression). 
Equation \eqref{eq:energyamplitude} is obtained when considering a set of vectors $\vec {q}$ of length $q_0$, while similar forms may be achieved when considering different length scales  \cite{elderAmplitudeExpansionBinary2010,salvalaglioMesoscaleDefectMotion2021}. 
The evolution of $\eta_n$, which follows from Eq.~\eqref{eq:PFC_diffusive_dynamics} is \cite{goldenfeldRenormalizationGroupApproach2005,salvalaglioCoarseGrainedModeling2022},
\begin{equation}
\frac{\partial \eta_n}{\partial t} =- \Gamma q_0^2 \frac{\delta F}{\delta \eta_n^*},
\label{eq:amplitudetime}
\end{equation}
with
\begin{equation}
\frac{\delta F}{\delta \eta_n^*}= \left[
\Delta B_0 + B_0^x\mathcal{G}_n^2 + 3V \left(\Phi-|\eta_n |^2\right)\right]\eta_n + \frac{\partial f^s}{\partial \eta_n^*},
\label{eq:amptimefuncder}
\end{equation}
 where the last term comes from the nonlinear contributions $\psi^3$ and $\psi^4$ in the local free energy density, and depend on the other amplitudes $\{ \eta_m  \}_{m\neq n}$.
 However, for the amplitudes that go to zero at the defect, it can be shown that $\frac{\partial f^s}{\partial \eta_n^*}=0$ at the defect (for more details, see \ref{appendix:amplitude_decoupling}). 
 Thus, the evolution of $\eta_n$ near the defect core is dictated solely by the non-local gradient term, namely
\[
\d_t \eta_n 
\approx -\Gamma B_0^xq_0^2\G_n^2 \eta_n.
\label{eq:amplitude_evolution_equation_simple}
\]
Furthermore, this implies that the complex amplitude $\eta_n$ of a stationary defect satisfy $\G_n^2\eta_n^{(0)}=0$ at the core. 
We now add an imposed, smooth displacement $\tilde{\vec u}$ to the amplitudes as $\eta_n = \eta_n^{(0)}e^{-i \vec q_n \cdot\tilde{\vec u}}$ to represent the far-field displacement induced by a different line segment, defect, or externally applied loads \cite{skaugenDislocationDynamicsCrystal2018}. 
This displacement is in addition to the discontinuous displacement field $\vec u$, described in Sec. \ref{sec:DD_VEL}, which is captured by stationary solution $\eta_n^{(0)}$ and defines the Burgers vector of the dislocation line (Fig.~\ref{fig:amplitude_discontinuity}). 
Inserting this ansatz of the complex amplitudes into Eq. \eqref{eq:determinant_vector}, and in the approximation of small distortions, $|\nabla \tilde{\vec u}|\ll 1$, we find
\[
\vec D^{(n)} =  \vec  D^{(n),0} + \frac{1}{2} \nabla (\vec  q^{(n)} \cdot \vec {\tilde u})  \times \nabla (|\eta_n^{(0)}|^2),
\qquad 
\left (
D_i^{(n)}(\vec r) = D_i^{(n),0} + \frac{1}{2} \epsilon_{irs} q_m^{(n)} (\partial_r \tilde u_m) \partial_s |\eta_n^{(0)}|^2
\right ),
\]
where $ D_i^{(n),0}$ is the determinant vector field calculated from $\eta_n^{(0)}$.
The corresponding defect density current is  
\[
\vec J^{(n)}
=
4 \Gamma B_0^x  q_0^2 \Im \left (\I (\nabla \eta_n^{(0)*}) \otimes (\nabla +\I \vec q^{(n)} )
 \mathcal G_n \eta_n^{(0)}\right ) \cdot \nabla (\vec q^{(n)}  \cdot \tilde {\vec u}) 
\qquad
\left (
J^{(n)}_l 
=
4 \Gamma B_0^x  q_0^2 q_i^{(n)}(\partial_k \tilde u_i)  \Im \left (\I (\partial_l \eta_n^{(0)*}) (\partial_k+\I q_k^{(n)} )
 \mathcal G_n \eta_n^{(0)}\right )
 \right ).
\]
Arguably, the simplest solution of Eq.~(\ref{eq:amplitude_evolution_equation_simple}) is the isotropic, simple vortex $\eta^{(0')}_n$ which is linear with the distance from the core and $s_n = \pm 1$. 
At a node $\vec r'$ on the dislocation line, $\eta^{(0')}$ can be written in terms of the Cartesian coordinates $x_\perp,y_\perp$ in the plane $\mathcal N'$ (Sec. \ref{sec:DD_VEL}), where it takes the form $\eta^{(0')}_n = \kappa ( x_\perp + \I s_n y_\perp)$, with $\kappa$ a proportionality constant.
The gradients of $\eta_n^{(0')}$ can be evaluated in these coordinates and gives at $\vec r'$, $\Im \left (\I(\d_l \eta_n^{(0')*})(\d_m \eta_n^{(0')})\right ) =\kappa^2 (\delta_{lm} - t'_l t'_m)$, from which we get the current
\[
\vec J^{(n)} 
=
-8 \kappa^2 \Gamma B_0^x  q_0^2  (\vec q^{(n)} \cdot \nabla(\vec q\cdot \tilde {\vec u}_i))  (\mathbbm 1 - \vec t'\otimes \vec t')\cdot \vec q^{(n)}
\qquad
\left (
J^{(n)}_l 
=
-8 \kappa^2 \Gamma B_0^x  q_0^2 q_i^{(n)}q_k^{(n)}q_m^{(n)} (\partial_k \tilde u_i)  (\delta_{lm} - t'_l t'_m)
\right )
\]
in terms of the local tangent vector $\vec t'$.
At $\vec r'$, we also get $D_i^{(n)} = \kappa^2 s_n t_i'$, which leads to an expression of the dislocation velocity (where the proportionality constant $\kappa$ cancels out), given by 
\[
v_s' = 
- \epsilon_{slr} \frac{\Gamma \pi}{|\vec b|^2}  b_j t_r' 4 B_0^x \sum_{n=1}^{12}       q_i^{(n)}q_j^{(n)} q_k^{(n)}q_l^{(n)} (\partial_k \tilde u_i)
= 
\frac{\Gamma \pi }{\eta_0^2 |\vec b|^2} \epsilon_{srl} t_r' \tilde \sigma_{lj} b_j, 
\]
where $\tilde \sigma_{lj}$ is the stress tensor for a bcc PFC that has been deformed by $\tilde{\mathbf{u}}$ \cite{skogvollStressOrderedSystems2021},
\[
\tilde \sigma_{lj}= 4  B_0^x \eta_0^2 \sum_{n=1}^{12} q_i^{(n)} q_j^{(n)} q_k^{(n)} q_l^{(n)}\d_{k}  \tilde u_{j}.
\]
Thus, the velocity of the dislocation line is proportional to the stress on the line. 
In vectorial form, this equation reads 
\[
\vec v
=
M
\vec t
\times 
(\tilde \sigma \cdot \vec b).
\label{eq:dislocation_overdamped_motion}
\]
with isotropic mobility $M = \Gamma \pi/(|\vec b|^2 \eta_0^2)$. 

A stationary dislocation induces a stress field $\sigma_{ij}^{(0)}$, but only the imposed stress $\tilde \sigma_{ij}$ appears in the equation above. 
This is analogous to how the stress field of the dislocation itself is not included when the Peach-Koehler force as calculated \cite{kosevichCrystalDislocationsTheory1979}. Thus, if $\sigma_{ij}^\psi$ is the configurational stress of the phase field at any given time, the part responsible for dislocation motion is the imposed stress 
\[
\tilde \sigma_{ij} = \sigma_{ij}^\psi - \sigma_{ij}^{(0)}.
\]
Note that the stationary solution necessarily satisfies mechanical equilibrium, $\partial_j \sigma_{ij}^{(0)}$, so that if the configurational PFC stress $\sigma_{ij}^\psi$ is in mechanical equilibrium, so is the imposed stress $\tilde \sigma_{ij}$ on the dislocation segment. 
The imposed stress used can be attributed to external load, other dislocations, or other parts of the dislocation loop. 
The framework predicts a defect mobility which is isotropic and does not discriminate between dislocation climb and glide motion.
Numerically however, we have seen that at deeper quenches $\Delta B_0$, climb motion is prohibited in the PFC model. 
The result in this section should therefore be interpreted as a first-order approximation, valid at shallow quenches.
This apparent equal mobility for glide and climb may result from the employment of the amplitude phase-field model (which is only exact for $|\Delta B_0| \rightarrow 0$) or the assumption of an isotropic defect core in the calculation.


\subsection{PFC dynamics constrained to mechanical equilibrium (PFC-MEq)}
\label{sec:PFC_mechanical_equilibrium}

In the previous section, we found that the motion of a dislocation is governed by a configurational stress $\sigma^\psi_{ij}$ which derives from the PFC free energy. 
Since this stress is a functional only of the phase field configuration, it does not satisfy, in general, the condition of mechanical equilibrium. 
References \cite{skaugenDislocationDynamicsCrystal2018, skogvollStressOrderedSystems2021} give an explicit expression for this stress defined as the variation of the free energy with respect to distortion, 
\[
\sigma^\psi_{ij} = -2B_0^x\avg{\L \psi \d_{ij} \psi},
\label{eq:stress_sigma}
\]
where $\avg{\cdot}$ is a spatial average over $1/q_{0}$ in order to eliminate the base periodicity of the phase field (see \ref{appendix:amplitude_demodulation}).

In this section, we discuss a modification of the PFC in three dimensions and in an anisotropic lattice so as to maintain elastic equilibrium in the medium while $\psi$ evolves according to Eq. (\ref{eq:PFC_diffusive_dynamics}). 
Let $\psi^{(U)}$ be the field that results from the evolution defined by Eq.  (\ref{eq:PFC_diffusive_dynamics}) alone. 
At each time, we define
\[
\psi(\vec r) = \psi^{(U)}(\vec r-\vec u^\delta),
\label{eq:mechanical_equilibrium_advection}
\]
where $\vec u^\delta$ is a small continuous displacement computed so that the configurational stress associated with $\psi(\vec r)$ is divergence free. 
We now show a method to determine $\vec u^\delta$. 
Suppose that at some time $t$ the PFC configuration $\psi$ has an associated configurational stress $\sigma_{ij}^{\psi,U}$ (from Eq. (\ref{eq:stress_sigma}), where $\partial_j\sigma_{ij}^{\psi,U} \neq 0$).
Within linear elasticity, the stress $\sigma_{ij}^{\psi}$ after displacement of the current configuration by $\vec u^\delta$ is given by 
\[
\sigma_{ij}^{\psi}
=
\sigma_{ij}^{\psi,U} + C_{ijkl} e^{\delta}_{kl},
\label{eq:mech_eq_central_assumption}
\]
where $C_{ijkl}$ is the elastic constant tensor, and $e_{ij}^\delta = \frac{1}{2} (\d_i u_j^\delta +\d_j u_i^\delta)$. 
$\vec u^\delta$ is determined by requiring that 
\[
\d_j \sigma_{ij}^{\psi}
=
\d_j (\sigma_{ij}^{\psi,U} + C_{ijkl} e^{\delta}_{kl} ) = 0.
\]
By using the symmetry $i\leftrightarrow j$ of the elastic constant tensor, we can rewrite this equation explicitly in terms of $\vec u^\delta$,
\[
g_i^{\psi,U} + C_{ijkl} \d_{jk} u_{l}^\delta = 0,
\label{eq:PDE_determining_u_delta}
\]
where 
\[
g_i^{\psi} = \d_j \sigma_{ij}^\psi = \avg{\frac{\delta F_\psi}{\delta \psi} \d_i \psi - \d_i  f}
\]
is the body force from the stress \cite{skogvollStressOrderedSystems2021}.
The quantity $f$ is the free energy density from Eq.~\eqref{eq:F_PFC}.

Given the periodic boundary conditions used, the system of equations (\ref{eq:PDE_determining_u_delta}) is solved by using a Fourier decomposition with the Green's function for elastic displacement in cubic anisotropic materials \cite{dederichsElasticGreenFunction1969}. 
Once $\vec u^{\delta}$ is obtained, $\psi$ is updated according to Eq.~\eqref{eq:mechanical_equilibrium_advection}, and evolved  according to Eq.~(\ref{eq:PFC_diffusive_dynamics}) from its current state $\psi(t)$ to $\psi^{(U)}(t+\Delta t)$.
Note that Eqs.~\eqref{eq:PDE_determining_u_delta} can, in general, be solved for any elastic constant tensor, so that the method introduced is not limited to cubic anisotropy. 
Since the state $\psi^{(U)}$ can only be updated according to Eq. \eqref{eq:mechanical_equilibrium_advection} every $\Delta t$, this effectively sets a time scale of elastic relaxation in the model. 
We found that the numerical discretization scheme for imposing mechanical equilibrium at every $\Delta t$ has a slow convergence with decreasing time resolution. 
Thus, the rate of loop shrinkage also depends slightly on $\Delta t$.
This is further discussed in \ref{appendix:numerical_methods}.

\begin{figure*}[htp]
    \centering
    \includegraphics[]{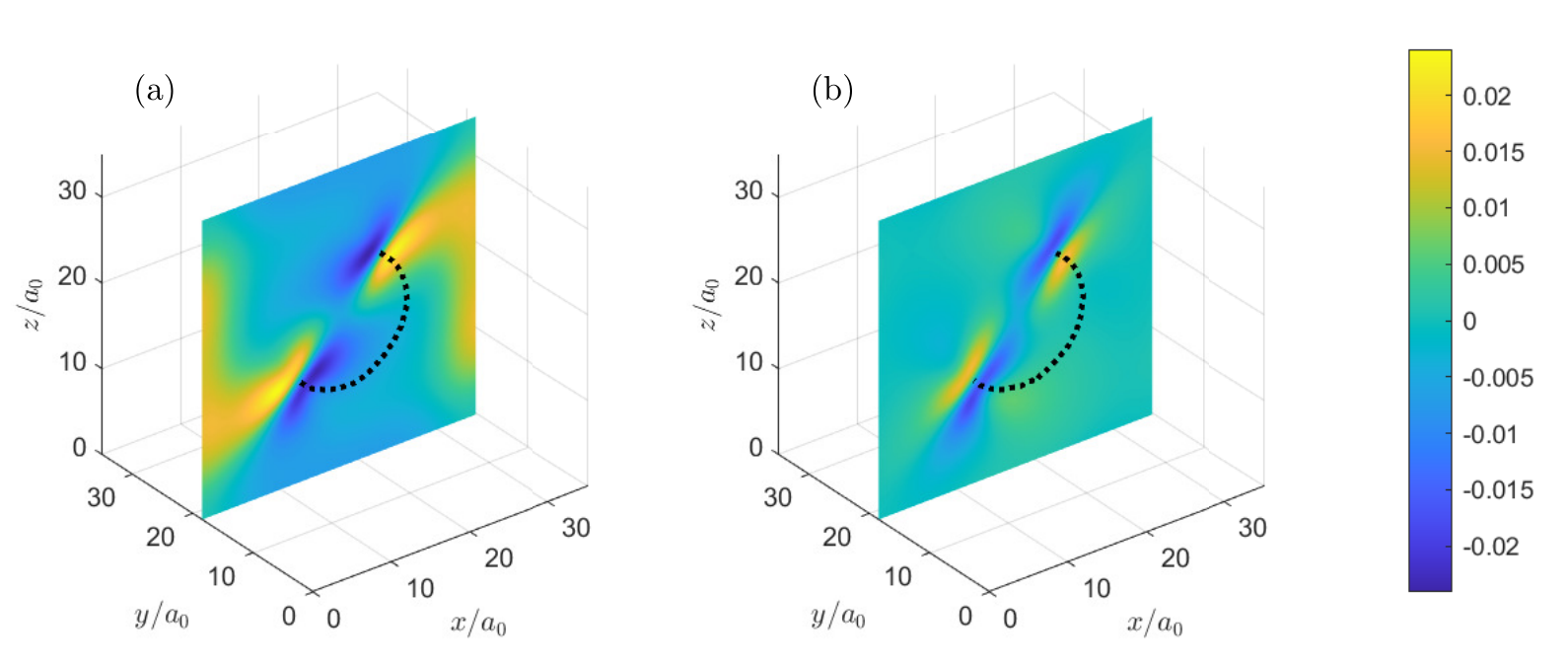}
    \caption{
    In-plane sections ($y=17.5a_0$) of the configurational stress $\sigma_{xz}^\psi/\mu$ for the dislocation loop after shrinking to $90\%$ of its initial circumference under (a) PFC dynamics and (b) PFC-MEq dynamics.
    Because the latter evolves faster, the snapshots are taken at different times, namely $t=389.0\tau$ and $t=34.4\tau$, respectively.
    A lot of residual (unrelaxed) stress is visible in the configurational stress for the classical PFC model.
    }
    \label{fig:loops_annealed_at_90_percent_3D_view}
\end{figure*}

Figure \ref{fig:loop_shrinkage} contrasts numerical results for the evolution of an initial dislocation loop with and without using the method just described. 
The computed line velocities are very different as they are highly sensitive to the local stress experienced by the dislocation loop segments. 
This stems from the fact that under classical PFC dynamics, the stress is always given by $\sigma_{ij}^{\psi,U}$, and a consequence of the results from Sec. \ref{sec:theoretical_PK_force} is that the velocity of an element of the defect line will be quite different depending on whether the stress acting on it is $\sigma_{ij}^{\psi,U}$ or $\sigma_{ij}^{\psi}$. 
Figure \ref{fig:loops_annealed_at_90_percent_3D_view} shows the dislocation loop after its circumference has shrunk to $90\%$ of its initial value, and the resulting $xz$ component of the stress for both models. 
As expected, the correction provided by the PFC-MEq model is necessary to relax the stress originating from the initial loop. 
The figure shows a large residual stress far from the dislocation loop that can only decay diffusively in the standard phase field model. 
Indeed, we have verified numerically that the configurational stress is only divergence-less for the PFC-MEq model. We note that in our set the loop is seeded in a glide plane, thus its shape remains approximately circular for both models, while the shrinkage rate is different. 
Note that with the addition of this advection step, the model is no longer guaranteed to be fully dissipative. 

The problem addressed in this section involves finding the elastic distortion $u_{kl}$ (which away from defects it can be written as $u_{kl} = \partial_k u_l$ for a displacement field $\vec u$) given the dislocation density tensor $\alpha_{ij}$ as a state variable \cite{acharyaStructureLinearDislocation2019}. The first part is the incompatibility of the elastic distortion
\[
\e_{ilm} \d_l u_{mk} =  - \alpha_{ik},
\label{eq:distortion_dislocation_density}
\]
and the second is the mechanical equilibrium condition on $u_{kl}$
\[
\d_j C_{ijkl} (u_{kl})_{S} = 0,
\label{eq:distortion_elastic_equilibrium}
\]
where $C_{ijkl}$ is the tensor of elastic constants, and $(S)$ denotes the symmetric part of the tensor. Equation \eqref{eq:distortion_dislocation_density} has a non trivial kernel consisting of gradients of vector fields $\nabla \vec u^\delta$. This vector field is determined by Eq. \eqref{eq:distortion_elastic_equilibrium} given appropriate boundary conditions that guarantee uniqueness. A computational method for solving for $u_{kl}$ and $\vec u^\delta$, using the dislocation density as a state variable, was first given in Ref. \cite{royFiniteElementApproximation2005}. The main difference between this reference and the method outlined in this section is that, since the incompatibility of the distortion is captured by the state of the phase field, we only need to solve for the compatible part of the distortion using the force density $\vec g^\psi$ from the phase field as a source. 

\begin{figure*}[htp]
    \centering
    \includegraphics[width=\textwidth]{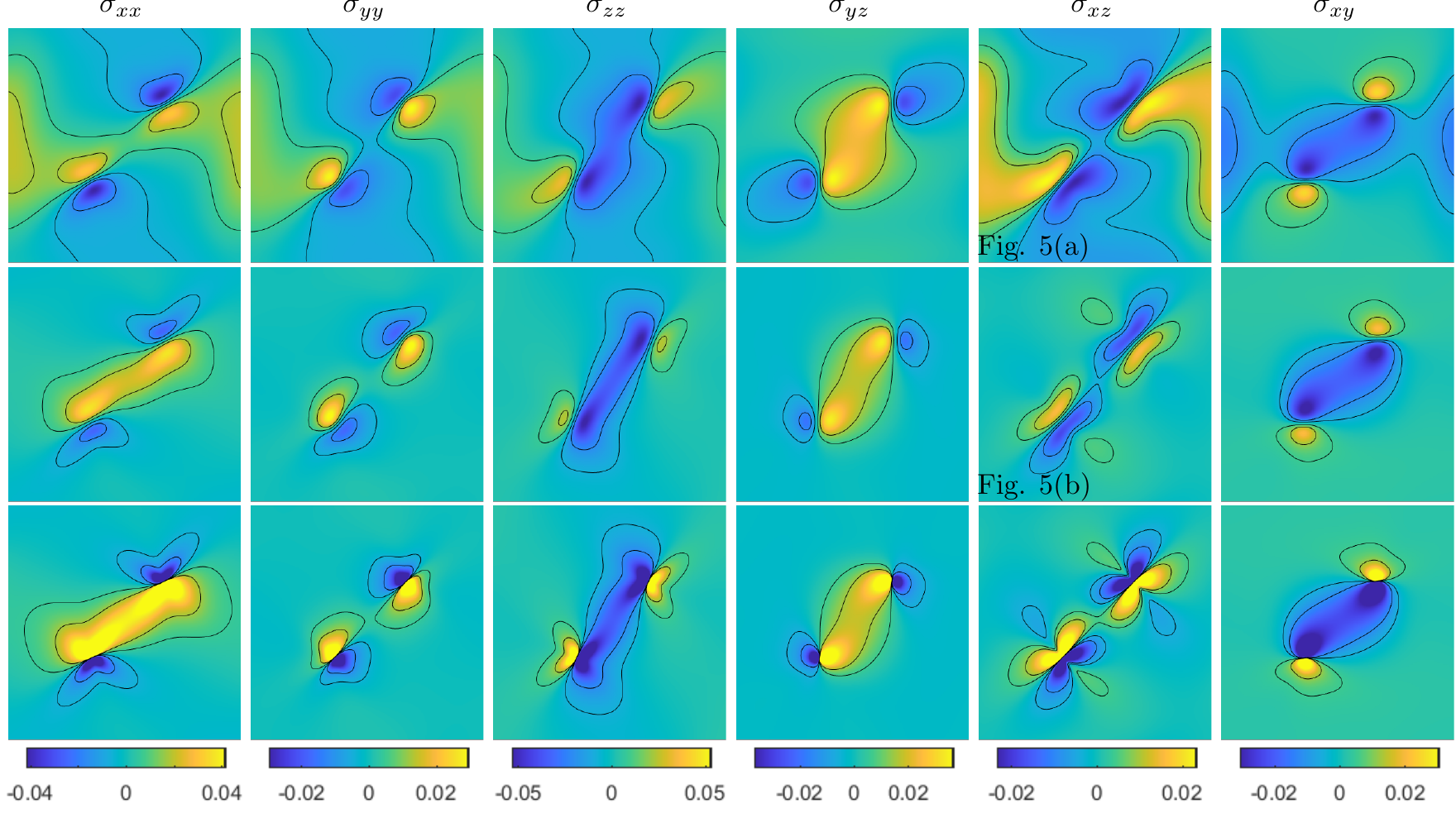}
    \caption{
    In-plane sections ($y=17.5 a_0$) of the stresses for the dislocation loop after it has shrunk to $90 \%$ of its initial circumference in (top row) the PFC model $\sigma_{ij}^{\psi,U}$, (middle row) the PFC-MEq model $\sigma_{ij}^\psi$ and (bottom row) the prediction from continuum elasticity $\sigma_{ij}^{(\alpha)}$ using the dislocation density extracted from the PFC as a source.
    The stresses predicted from continuum elasticity are singular, so the colorbar for each column is saturated at $\pm \max(|\sigma_{ij}^{\psi,U}|)$ and contour lines are drawn at $\pm 15\%,\pm 40\%$ of this value. 
    For the comparison, we have subtracted from $\sigma_{ij}^{\psi,U}$ and $\sigma_{ij}^\psi$ their mean values (see text).
    The stresses are given in units of the shear modulus $\mu$.}
    \label{fig:all_stresses_of_shrinkage_loops}
\end{figure*}

While the stress profile shown in Fig.~\ref{fig:loops_annealed_at_90_percent_3D_view}(b), can be shown numerically to have vanishing divergence, we would like to see a direct comparison of the stress with the prediction from continuum elasticity.
As the model purports to evolve the phase-field at mechanical equilibrium, and we are able to extract the dislocation density from the phase-field at any time through Eq. \eqref{eq:dislocation_density}, this amounts to the problem of finding the stress tensor for a given dislocation density, under the constraint of mechanical equilibrium and with periodic boundary conditions (zero surface traction). 
This problem was adressed in Ref. \cite{brennerNumericalImplementationStatic2014}, and in \ref{appendix:equilibrium_elastic_fields}, we show how we solve Equations (\ref{eq:distortion_dislocation_density}-\ref{eq:distortion_elastic_equilibrium}) to derive the equilibrium stress field from $\alpha_{ij}$ using spectral methods. 
Figure \ref{fig:all_stresses_of_shrinkage_loops} shows all the stress components after the dislocation loop has shrunk to $90\%$ of its initial diameter for both dynamical models, as well as the stress $\sigma^{(\alpha)}$ computed directly from the dislocation density tensor.\footnote{Due to the geometric similarity in how the loop annihilates in the different models, there is no observable difference in the continuum elastic stress field predictions between using $\alpha$ from either model as a source.}
Note that the mean value of the components of $\sigma_{ij}^{(\alpha)}$, is not determined by Eqs.~\eqref{eq:distortion_dislocation_density} - \eqref{eq:distortion_elastic_equilibrium}, and is set to zero. 
In this comparison, we have also subtracted from $\sigma_{ij}^\psi$ its mean value. 
As expected, the stresses obtained from the PFC-MEq model agree well with $\sigma_{ij}^{(\alpha)}$. 
The small differences observed are due to the fact that the configurational stress determined by $\psi$ is naturally regularized by the lattice spacing and the finite defect core, whereas the stress
$\sigma_{ij}^{(\alpha)}$ is for a continuum elastic medium with a singular dislocation source (numerically, the $\delta$-functions in Eq. \eqref{eq:dislocation_density} is regularized by an arbitrary width of the Gaussian approximation). 
Investigating exactly which length scale of core regularization derives from the PFC model is an open and interesting question that we will address in the future.

\section{Conclusions}

We have introduced a theoretical method, and the associated numerical implementation, to study topological defect motion in a three dimensional, anisotropic, crystalline PFC lattice. 
The dislocation density tensor and velocity are directly defined by the spatially periodic phase field, where dislocations are identified with the zeros of its complex amplitudes.

To illustrate the method, we have studied the motion of a shear dislocation loop, and found that it accurately tracks the loop position, circumference, and velocity. 
As an application, we have shown that under certain simplifying assumptions, the overdamped dislocation velocity follows from the Peach-Koehler force, with the defect mobility determined by equilibrium lattice properties. 
We have introduced the PFC-MEq model for three dimensional anisotropic media which constrains the classical PFC model evolution to remain in mechanical equilibrium, and shown that loop motion is much faster with this modification. 
The PFC-MEq model produces stress profiles that are in agreement, especially far from the defect core, to stress fields directly computed from the instantaneous dislocation density tensor.

In summary, we have presented a comprehensive framework, based on the phase field crystal model for the analysis of dislocation motion in crystalline phases in three spatial dimensions.
Starting from a free energy that has a ground state of the proper symmetry, the model naturally incorporates defects, the associated topological densities, and the resulting defect line kinematic laws that are compatible with topological density conservation. 
Configurational stresses induced by defects are defined and analyzed, and shown to lead to a Peach-Koehler type force on defects, with an explicit expression for the line segment mobility given.

\section*{Acknowledgements}
V.S. and L.A. acknowledge support from the Research Council of Norway
through the Center of Excellence funding scheme, Project No. 262644 (PoreLab). 
M.S. acknowledges support from the Emmy Noether Programme of the German Research Foundation (DFG) under Grant No. SA4032/2-1. The research of J.V. is supported by the National Science Foundation, contract No. DMR-1838977.


\appendix

\section{Numerical methods}
\label{appendix:numerical_methods}


\subsection{Amplitude demodulation}
\label{appendix:amplitude_demodulation}

Given a phase-field configuration described by slowly varying amplitudes $\eta_n(\vec r)$ 
\[
\psi(\vec r) = \bar \psi (\vec r) + \sum_{n'} \eta_{n'}(\vec r) e^{i\vec q^{(n')} \cdot \vec r},
\label{eq:PFC_by_coarse_amplitudes}
\]
we can find the amplitudes using the principle of resonance under coarse graining.
Coarse graining $ \tilde X $ with respect to a length scale $a_0$ is introduced as a convolution with a Gaussian filter function 
\[
\langle {\tilde X} \rangle (\vec r) = \int d\vec r'\frac{\tilde X(\vec r')}{(2\pi a_0^2)^{d/2}} \exp\left (-\frac{(\vec r-\vec r')^2}{2a_0^2}\right).
\]
Given the PFC configuration of Eq. \eqref{eq:PFC_by_coarse_amplitudes}, to find $\eta_n(\vec r)$, we multiply by $e^{-\vec q^{(n)} \cdot \vec r}$ and coarse grain to get
\[
\left \langle  \psi(\vec r) e^{-\vec q^{(n)} \cdot \vec r}
\right \rangle 
=
\bar \psi (\vec r) \left \langle  e^{-\vec q^{(n)} \cdot \vec r} \right  \rangle  + \sum_{n'} \eta_{n'}(\vec r)\left  \langle e^{i(\vec q^{(n')} - \vec q^{(n)}) \cdot \vec r} \right \rangle 
=
\eta_n(\vec r)
,
\]
where we have used the slowly varying nature of the complex amplitudes to pull them out of the coarse graining operation and used the resonance condition $\langle e^{i(\vec q^{(n')} - \vec q^{(n)})\cdot \vec r}\rangle = \delta_{nn'}$ \cite{skogvollStressOrderedSystems2021}.

\subsection{Dislocation density tensor decomposition}
\label{appendix:SVD_of_alpha_in_tiBj}

A singular value decomposition of $\alpha$ is introduced as $\alpha = U \Sigma V^T$, where $\Sigma$ is a diagonal matrix containing the singular values of $\alpha$, and $U$ and $V$ are unitary matrices containing the normalized eigenvectors of $(\alpha \alpha^T)$ and $(\alpha^T \alpha)$, respectively. 
We assume that the dislocation density tensor can be written as the outer product of the unitary tangent vector $\vec t$ and a local spatial Burgers vector density $\vec B(\vec r)$, i.e., $\alpha_{ij} = t_i B_j$. 
Under this assumption, one finds $\Sigma$ with only one non zero singular value,  $|\vec B|$, and the columns of $U$ and $V$ that correspond to this singular value will be $\vec t$ and ${\vec B}/{|\vec B|}$, respectively.

\subsection{Evolution of the phase field}
\label{appendix:numerical_methods_phase_field_evolution}

The dimensionless parameters for the bcc ground state are set to: $\Delta B_0=-0.3, T=0$ and $\psi_0= -0.325$. 
Lengths have been made dimensionless by choosing $|\vec q^{(n)}| = q_0 = 1$, yielding a bcc lattice constant $a_0 = 2\pi \sqrt{2}$. 
In all simulations, the computational domain is given by $35\times35\times35$ base periods of the undistorted bcc lattice, with grid spacing $\Delta x = \Delta y = \Delta z = a_0/7$. 
Periodic boundary conditions are used throughout. 
Equation (\ref{eq:PFC_diffusive_dynamics}) is integrated forward in time with an explicit method \cite{coxExponentialTimeDifferencing2002}, and $\Delta t = 0.1$.  
A Fourier decomposition of the spatial fields is introduced to compute the spatial derivatives of the fields, while nonlinear terms are computed in real space.

\subsubsection{Mechanical equilibrium}

We implement the correction scheme of Eq.~\eqref{eq:mechanical_equilibrium_advection} between every timestep $\Delta t$. 
If $u_{\max} = \max_{\vec r\in \textrm{Domain}} (\vec u^\delta(\vec r))>0.1a_0$, we rescale $\vec u^\delta$ so that $u_{\max}=0.1a_0$, and repeat the process again until elastic equilibrium is achieved. 
Typically, when initializing the PFC field with a dislocation, around $5$ such iterations are needed, after which, $u_{\max}$ is on the order of $0.01 a_0$ at each correction step.

The dislocation loop shrink velocity is sensitive to the time interval $\Delta t$ between each equilibration correction.
As shown in Fig.~\ref{fig:loop_shrinkage}, the effect of imposing this correction at every time interval $\Delta t=0.1$ accelerates the annihilation process by approximately a factor of $|\vec v_{\textrm{PFCMEq}, \Delta t=0.1}|/|\vec v_{\textrm{PFC}}|\approx7.5$.
A slow convergence  in the limit $\Delta t\rightarrow 0$ is observed, where we have estimated that the shrink velocity increases up to $|\vec v_{\textrm{PFCMEq},\Delta t\rightarrow 0}|/|\vec v_{\textrm{PFC}}|\approx 9.8$. 
However, to reach this numerical convergence is computationally demanding. 
Indeed, this slow convergence suggests that the time scale of the elastic field relaxation is important for the process of shear dislocation loop shrinkage. 
For static problems however, such as obtaining regularized stress profiles for dislocation loops, or defect nucleation under quasi-static loading, this slow convergence is not an issue.

\subsection{Initializing a dislocation loop in the PFC model}
\label{appendix:initial_condition_explanation}
\begin{figure}
    \centering
    \includegraphics{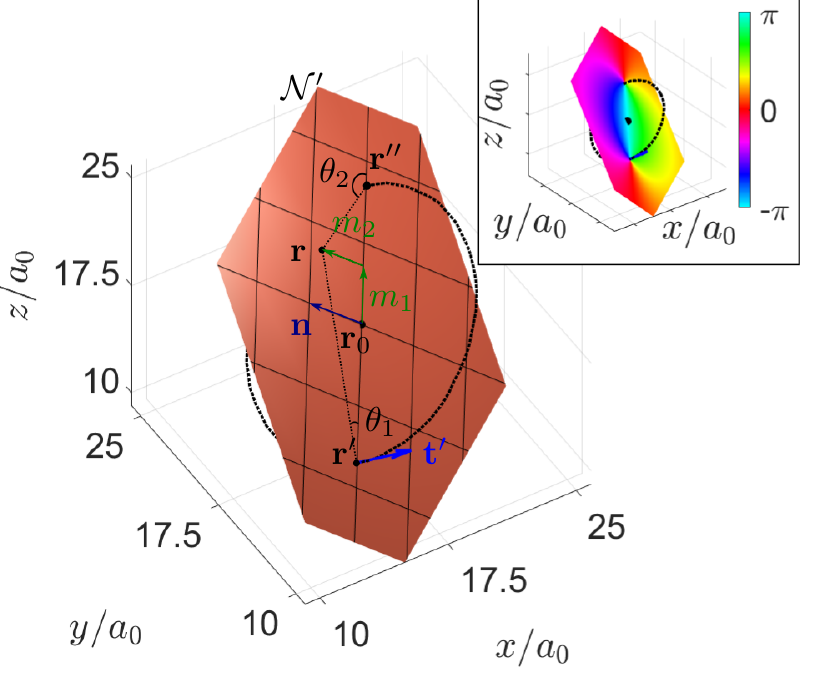}
    \caption{Geometry of the circular dislocation loop in a slip plane given by the normal vector $\vec n$. 
    $\mathcal N'$ is the plane normal to the tangent vector $\vec t'$ upon which we impose a Cartesian coordinate system to determine the angles $\theta_1$, $\theta_2$ that are used to construct the (inset) initial amplitude phase configuration. 
    For more details, see Section \ref{appendix:initial_condition_explanation}.}
    \label{fig:initial_condition_explanation}
\end{figure}

In this section, we show how to multiply the initial amplitudes $\eta_0$ with complex phases, to produce a dislocation loop with Burgers vector $\vec b$ in a slip plane given by normal vector $\vec n$ (see Sec. \ref{sec:numerical_loop_shrinkage}). 
Given a point $\vec r$, it belongs to a plane $\mathcal N'$ perpendicular to $\vec t'$ for some point $\vec r'$ on the dislocation loop (see Figure \ref{fig:initial_condition_explanation}).
This plane also intersects the diametrically opposed point $\vec r''$ of the dislocation loop. 
If $\vec r_0$ is the center of the loop, the distance vector $\vec r-\vec r_0$ lies in $\mathcal N'$. 
Let $(m_1,m_2)$ be the first and second coordinate in the Cartesian coordinate system defined by the right-handed orthonormal system $\{(\vec n \times \vec t'),\vec n,\vec t'\}$ centered at $\vec r_0$.
If $m_1>0$, we get from geometrical considerations
\[
m_2 = (\vec r-\vec r_0)\cdot \vec n,
\]
\[
m_1 = |(\vec r-\vec r_0) - m_2\vec n|.
\]
Both $m_1$ and $m_2$ are thus determined by $\vec r$, $\vec r_0$ and the normal vector to the loop plane $\vec n$.
$\theta_1$ ($\theta_2$) is the angle between $\vec r-\vec r'$ ($\vec r-\vec r''$) and $\vec n\times \vec t'$ in the plane $\mathcal N'$ and are found numerically by using the four-quadrant inverse tangent $\textrm{atan2}(y,x)$, so that
\[
\theta_1 = \textrm{atan2}\left (m_2,m_1+R\right ) 
\label{eq:theta1}
\]
\[
\theta_2 = \textrm{atan2}\left (m_2,m_1-R\right ) ,
\label{eq:theta2}
\]
where $R$ is the radius of the loop.
For each point $\vec r$, we determine $\theta_1 (\vec r)$ and $\theta_2(\vec r)$ according to the equations above and initiate the PFC with the phases
\[
\eta_n = \eta_0 e^{\mathbbm is_n (\theta_1(\vec r) - \theta_2(\vec r))},
\]
where $s_n = \frac{1}{2\pi} \vec q^{(n)} \cdot \vec b$ is given in Table \ref{tab:dislocharges}.
\renewcommand{\arraystretch}{1.2}
\begin{table}[h]
\centering
\begin{tabular}{ccccccc}
\toprule
$\vec b$  &  $s_1$ &  $s_2$ &  $s_3$ &  $s_4$ &  $s_5$ &  $s_6$  \\
\midrule
$\frac{a_0}{2} (-1,1,1)$     & 1&0&0&0&1&1 \\
$\frac{a_0}{2} (1,-1,1)$     & 0&1&0&1&0&-1\\
$\frac{a_0}{2} (1,1,-1)$     & 0&0&1&-1&-1&0\\
$\frac{a_0}{2} (1,1,1)$     &1&1&1&0&0&0 \\
$a_0 (1,0,0)$     &0&1&1&0&-1&-1 \\
$a_0 (0,1,0)$     &1&0&1&-1&0&1 \\
$a_0 (0,0,1)$     & 1&1&0&1&1&0\\
\bottomrule
\end{tabular}
\caption{Dislocation charges $s_n = \frac{1}{2\pi} \vec b\cdot \vec q^{(n)}$ for different Burgers vectors $\vec b$ in the bcc lattice.
 $\vec q^{(n)}$ is defined in Eq.~\eqref{eq:qn_bcc}.
}
\label{tab:dislocharges}
\end{table}
This ensures that the complex phases have the right topological charge (Eq. \eqref{eq:s_n}).
The inset in Fig.~\ref{fig:initial_condition_explanation} shows the phase of $e^{\mathbbm i (\theta_1-\theta_2)}$ in $\mathcal N'$ for $\vec n=\frac{1}{\sqrt 2} [-1,0,1]$, which is the slip plane chosen in the simulation in Sec. \ref{sec:numerical_loop_shrinkage}.
Note that points $\vec r$ for which $m_1<0$ are computed by the same equation for the tangent vector at $\vec r''$, with the same formula, thus validating the Eqs.~\eqref{eq:theta1}--\eqref{eq:theta2} for all values in $\mathcal N'$. 
Since the expressions are independent of the particular plane $\mathcal N'$ and each point $\vec r$ belongs to one such plane, they are also valid for all points in the simulation domain. 

\subsection{Calculating the perimeter of a dislocation loop}
\label{appendix:numerical_methods_dislocation_perimeter}

To calculate numerically the perimeter of a dislocation loop, recall that 
\[
\alpha_{ij} =b_j \int_{\C} dl_i' \delta^{(3)} (\vec r-\vec r_{(l)}' ),
\]
where we have added a subscript $(l)$ onto $\vec r'$ to emphasize that it is the point on the loop as indexed by the line element $d\vec l$. 
Taking the double dot product with itself, we find
\[
\alpha_{ij} \alpha_{ij}
=
|\vec b|^2 
\int_{\C} \int_{\C} 
dl_i' 
dm_i'
\delta^{(3)} (\vec r-\vec r_{(l)}' )
\delta^{(3)} (\vec r-\vec r_{(m)}' ).
\]
The contributions to this integral will only come from points on the loop $\C$ and only when $\vec r_{(l)}' = \vec r_{(m)}'$, where $dl_i' =
dm_i'$, so $d l_i d m_i = (d \vec l)^2 = |d l_i|^2 = |d l_i| |d m_i|$.
Thus
\begin{equation}
\alpha_{ij} \alpha_{ij}
=
|\vec b|^2 
\int_{\C} 
|dl_i'| 
\delta^{(3)} (\vec r-\vec r_{(l)}' )
\int_{\C} 
|dm_i'|
\delta^{(3)} (\vec r-\vec r_{(m)}' ) 
=
|\vec b|^2 
\left (
\int_{\C} 
|dl_i'| 
\delta^{(3)} (\vec r-\vec r_{(l)}' )
\right )^2.
\end{equation}
Taking the square root and integrating over all space, we find 
\[
\int d^3 r
\sqrt{
\alpha_{ij} \alpha_{ij}
}
=
|\vec b|
\int_C |d l_i|
\int d^3 r
\delta^{(3)} (\vec r-\vec r'_{(l)})
=
 |\vec b| L,
\]
where $L$ is the perimeter of the dislocation loop. Thus, 
\[
L = \frac{1}{|\vec b|} \int d^3 r \sqrt{
\alpha_{ij} \alpha_{ij}
}.
\]

\subsection{Direct computation of stress fields}
\label{appendix:equilibrium_elastic_fields}

The dislocation density tensor is calculated directly from the phase field $\psi$ through Eq.~\eqref{eq:dislocation_density}. 
The general method of solving Eqs. (\ref{eq:distortion_dislocation_density}-\ref{eq:distortion_elastic_equilibrium}) on a periodic medium is given in Ref. \cite{brennerNumericalImplementationStatic2014} given $\alpha_{ij}$, where also the uniqueness of the elastic fields is proven given appropriate conditions on the dislocation density $\alpha_{ij}$.
In the present case, the conditions on $\alpha_{ij}$ are automatically satisfied as it is calculated from the phase-field. 
In this section, we thus show for our computational setup, how we compute the Green's function in the relating the distortion $u_{ij}$ to the dislocation density tensor $\alpha_{ij}$ as a source.
Since (\ref{eq:distortion_dislocation_density}-\ref{eq:distortion_elastic_equilibrium}) given the periodic boundary conditions can be solved uniquely, we Fourier transform both sets of equations and add the condition of mechanical equilibrium (Eq. \eqref{eq:distortion_elastic_equilibrium}) to the diagonal equations ($i=k$) in  Eq.~(\ref{eq:distortion_dislocation_density}), which gives in Fourier space
\[
 \delta_{(i)k}  \frac{\I C_{(i)jml}}{\mu}  q_j \tilde u_{ml} 
 - \I\e_{ilm}  q_l \tilde u_{mk}  =  \tilde \alpha_{ik}, 
 \label{eq:distortion_in_fourier_space_full_set}
\]
where there is no summation over $(i)$, and we have multiplied the elastic constant tensor by $\I / \mu$ where $\mu$ is the shear modulus of the cubic lattice, and $C_{ijkl} = \lambda \delta_{ij} \delta_{kl} + \mu (\delta_{ik} \delta_{jl} + \delta_{il} \delta_{jk}) + \gamma \delta_{ijkl}$. By defining the 1D vectors $\vec{\tilde U}$ and $\vec{\tilde  \alpha}$ as
\begin{eqnarray*}
\vec {\tilde U}^T &= 
\begin{pmatrix}
u_{11}, & u_{12}, &
u_{13}, &
u_{21}, &
u_{22}, &
u_{23}, &
u_{31}, &
u_{32}, &
u_{33}
\end{pmatrix},
\\ 
 \vec{\tilde \alpha}^T&= 
\begin{pmatrix}
\alpha_{11}, &
\alpha_{12}, &
\alpha_{13}, &
\alpha_{21}, &
\alpha_{22}, &
\alpha_{23}, &
\alpha_{31}, &
\alpha_{32}, &
\alpha_{33}
\end{pmatrix},
\end{eqnarray*}
we rewrite Eq.~(\ref{eq:distortion_in_fourier_space_full_set}) more compactly as 
\[
M (\vec q) \vec {\tilde U} =  \vec{\tilde \alpha},
\]
where the explicit form of $M(\vec q)$ in the case of cubic anisotropy is given by
%
\[
M (\vec q)
=
\I \begin{pmatrix}
\frac{\lambda + 2\mu + \gamma}{\mu} q_1 & q_2       & q_3       & q_2 + q_3    & \frac{\lambda}{\mu} q_1                   & 0              & -q_2 + q_3    & 0         & \frac{\lambda}{\mu} q_1 \\
0                                       & 0         & 0         & 0             & q_3                                       & 0             & 0             & -q_2      & 0 \\
0                                       & 0         & 0         & 0             & 0                                         & q_3           & 0             & 0         & -q_2 \\
-q_3                                    &0          & 0         & 0             & 0                                         & 0             & q_1           & 0         & 0 \\
\frac{\lambda}{\mu} q_2                 &q_1-q_3    & 0         & q_1           & \frac{\lambda + 2\mu + \gamma}{\mu} q_2   & q_3           & 0             & q_1 + q_3 & \frac{\lambda}{\mu} q_2 \\
0                                       & 0         &-q_3       & 0             & 0                                         & 0             & 0             & 0         & q_1 \\
q_2                                     & 0         & 0         & -q_1          & 0                                         & 0             & 0             & 0         & 0 \\
0                                       & q_2       & 0         & 0             & -q_1                                      & 0             & 0             & 0         & 0 \\
\frac{\lambda}{\mu} q_3                 & 0         &q_1 + q_2  & 0             & \frac{\lambda}{\mu} q_3                   & -q_1 + q_2    & q_1           & q_2       & \frac{\lambda + 2\mu + \gamma}{\mu} q_3\\ 
\end{pmatrix}.
\]
%
$M(\vec q)$ can be inverted to yield the Fourier transform of the distortion $\vec U$,
\[
\tilde {\vec U} = M^{-1} (\vec q) \vec {\tilde  \alpha}.
\]
Once $\vec{\tilde U}$ (denoted by $\tilde u_{kl}$ in components) is known, we compute the stress field in mechanical equilibrium 
\[
\tilde \sigma_{ij} = C_{ijkl} \tilde u_{kl}.
\]
The dislocation density $\alpha_{ik}$ as obtained from the phase field as in Eq.~\eqref{eq:dislocation_density} has a very small divergence due to numerical round-off errors. 
We impose $\partial_i \alpha_{ik} = 0$ explicitly before evaluating $\tilde \sigma$, which improves numerical stability.

\section{Inversion formula for highly symmetric lattice vector sets}
\label{appendix:inversion_identity}

In inverting Eq.~(\ref{eq:thetan}) to obtain the displacement field $\vec u$ in terms of the phases $\theta_n$, we used the result of Eq.~(\ref{eq:Qij_identity}).  
This follows from the properties of moment tensors constructed from lattice vector sets $\mathcal Q = \{\vec q^{(n)} \}_{n=1}^N$. The $p$-th order moment tensor constructed from $\mathcal Q$ is given by 
\[
Q_{i_1...i_p} = \sum_{n=1}^N q_{i_1}^{(n)}  ...  q_{i_p}^{(n)}.
\]
 In two dimensions, for a parity-invariant lattice vector set that has a B-fold symmetry, Ref.~\cite{chenMomentIsotropyDiscrete2011} showed that all $p$-th order moments vanish for odd $p$ and are isotropic for $p<B$. Every isotropic rank 2 tensor is proportional to the identity tensor $\delta_{ij}$, so for a 2D lattice vector set having four-fold symmetry, such as the set of shortest reciprocal lattice vectors $\{\vec q^{(n)}\}_{n=1}^4$  of the square lattice, we have $\sum_{n=1}^4 q_i^{(n)} q_j^{(n)} \propto \delta_{ij}$ (Figs. 3 and 5 in Ref.~\cite{skogvollStressOrderedSystems2021} show the reciprocal lattice vector sets discussed in this appendix).  
 Taking the trace and using that the vectors have the same length $|\vec q^{(n)}|=q_0$, we get $\sum_{n=1}^4 q_i^{(n)} q_j^{(n)} = 4q_0^2\delta_{ij}$. 
 In general, for any 2D parity invariant lattice vector set $\{\vec q^{(n)}\}_{n=1}^N$  with a $B$-fold symmetry where $B>2$, we have 
\[
\textrm{2D:}\quad \sum_{n=1}^N q_i^{(n)} q_{j}^{(n)} = \frac{Nq_0^2}{2}\delta_{ij}. 
\]
As mentioned, this holds for the 2D square lattice, but it also holds for the 2D hexagonal lattice. 
In fact, the six-fold symmetry of the hexagonal lattice ensures that also every fourth-order moment tensor is isotropic, which results in elastic properties of the 2D hexagonal PFC model being isotropic \cite{skogvollStressOrderedSystems2021}. 

To show this identity for a 3D parity invariant vector set with cubic symmetry, we generalize the proof in Ref.~\cite{chenMomentIsotropyDiscrete2011} to a particular case of a 3D vector set that is symmetric with respect to $90^\circ$ rotations around each coordinate axis, such as the set of shortest reciprocal lattice vectors $\{\vec q^{(n)}\}_{n=1}^N$ of bcc, fcc or simple cubic structures. 
Let $\vec v$ be an eigenvector of $Q_{ij}$ with eigenvalue $\lambda$, i.e., $Q_{ij} v_j = \lambda v_i$. 
Since $Q_{ij}$ is invariant under a $90^\circ$ rotation $R^{(x)}_{ij}$ around the $x$-axis (i.e., $R^{(x)}_{ik} Q_{kl} ({R^{(x)}}^T)_{kj} = Q_{ij}$), we get $Q_{ij} R_{jl}^{(x)} v_l = \lambda R_{il}^{(x)} v_l$, showing that $R^{(x)} \vec v$ is also an eigenvector of $Q_{ij}$ with the same eigenvalue $\lambda$. 
Repeating for a rotation around the $y$-axis demonstrates that $Q_{ij}$ has only one eigenvalue $\lambda$, so that it must be proportional to the rank 2 identity tensor $Q_{ij} \propto \delta_{ij}$. 
Taking the trace and using that the vectors have the same length $|\vec q^{(n)}|=q_0$, we find
\[
\textrm{3D:}\quad \sum_{n=1}^N q_i^{(n)} q_{j}^{(n)} = \frac{Nq_0^2}{3}\delta_{ij}. 
\]

\section{Time derivatives of the dislocation density tensor}

\subsection{Delta-function form}
\label{appendix:derivative_of_dislocation_density_tensor_delta_function_form}
Consider a moving dislocation line $\mathcal C = \{\vec r'(\lambda,t)\}$ of points $\vec r(\lambda,t)$ parametrized by the time $t$ and a dimensionless $\lambda$ which can be taken to go from $0$ to $1$ without loss of generality. 
Keeping the labelling fixed through its time evolution, we get
\[
\alpha_{ij} (\vec r,t)   = b_j \int_{\lambda=0}^1  \delta^{(3)}(\vec r-\vec r'(\lambda,t)) (\partial_\lambda r_i'(\lambda,t)) d\lambda.
\]
Suppressing the dependence of $\vec r'$ on $\lambda$ and $t$, we get taking the time derivative of Eq. \eqref{eq:alpha_ik_definition_in_terms_of_ti_bk}, 
\[
\partial_t \alpha_{ij} = 
\underset{(1)}{\underbrace{
b_j \int_{\lambda=0}^1  (\partial_t \delta^{(3)}(\vec r-\vec r')) (\partial_\lambda r_i' ) d\lambda
}}
+
 \underset{(2)}{\underbrace{
 b_j \int_{\lambda=0}^1  \delta^{(3)}(\vec r-\vec r') 
 (\partial_t \partial_\lambda r_i' ) d\lambda
 }}.
\]
Starting with the first term using the chain rule, we have 
\[
(1)=
b_j \int_{\lambda=0}^1  (\partial_{k'} \delta^{(3)}(\vec r-\vec r')) V_k(\vec r')  (\partial_\lambda r_i' ) d\lambda,
\]
where $\partial_{k'} = \partial / \partial r_k'$ and $V_{k}$ is a field at time $t$ which is defined on $\vec r'\in \mathcal C$ as $\vec V(\vec r') = \partial_t \vec r'$, the velocity of the line segment perpendicular to the tangent vector.
We can rewrite $\partial_{k'} \delta^{(3)}(\vec r-\vec r') =-\partial_{k} \delta^{(3)}(\vec r-\vec r') $ and pull it outside the integral. 
Additionally, since $V_k(\vec r')$ is multiplied by a delta function, we can replace it by $V_k(\vec r)$, so we get
\[
(1)=
- \partial_k \left(
(
b_j \int_{\lambda=0}^1  \delta^{(3)}(\vec r-\vec r')  (\partial_\lambda r_i' ) d\lambda)(V_k(\vec r) )
\right )
= -\partial_k (\alpha_{ij} V_k)
.
\]
Turning to the second term, we get
\[
(2) = b_j \int_{\lambda=0}^1  \delta^{(3)}(\vec r-\vec r') 
 ( \partial_\lambda  V_i(\vec r')) d\lambda
 =
  b_j \int_{\lambda=0}^1  \delta^{(3)}(\vec r-\vec r') 
 ( \partial_{k'}V_i(\vec r'))  (\partial_\lambda r_k') d\lambda.
\]
Since $\partial_{k'}V_i(\vec r')$ is multiplied with a delta-function inside the integral, we can replace it by $\partial_{k}V_i(\vec r)$.
We thus get
\[
(2) = 
\left (
b_j \int_{\lambda=0}^1  \delta^{(3)}(\vec r-\vec r')   (\partial_\lambda r_k') d\lambda
\right )(  \partial_{k}V_i)
=
\alpha_{kj} \partial_k V_i =\partial_k( \alpha_{kj}  V_i),
\]
since $\partial_k \alpha_{kj} = \partial_k 
( b_j \int_{\mathcal C} \delta^{(3)}(\vec r-\vec r') d r'_k) = -b_j \int_{\mathcal C} (\partial_{k'} \delta^{(3)}(\vec r-\vec r')) d r'_k = -b_j [\delta(\vec r-\vec r')]_{\vec r'(\lambda=0)}^{\vec r'(\lambda=1)} 
=0$, either because $\mathcal C$ is a loop such that $\vec r'(\lambda=0) = \vec r'(\lambda=1)$ or else $\vec r'(\lambda=0)\neq\vec r\neq\vec r'(\lambda=1)$ since the dislocation cannot end inside the crystal.
This gives 
\[
\partial_t \alpha_{ij} =-\partial_k (\alpha_{ij} V_k)+ \partial_k( \alpha_{kj}  V_i)
= - \epsilon_{ikl} \partial_k (\epsilon_{lmn} \alpha_{mj} V_n).
\]

\subsection{Amplitude form}
\label{appendix:derivative_of_dislocation_density_tensor_amplitude_form}

Taking the time derivative of Eq. \eqref{eq:dislocation_density}, we have
\[
\partial_t \alpha_{ij} = 
\underset{(1)}{\underbrace{
\frac{6 \pi}{N q_0^2} \sum_{n=1}^N q_{j}^{(n)}  (\partial_t D^{(n)}_i) \delta^{(2)}(\eta_n)
}}
+
\underset{(2)}{\underbrace{
\frac{6 \pi}{N q_0^2} \sum_{n=1}^N q_{j}^{(n)}  D^{(n)}_i \partial_t \delta^{(2)}(\eta_n)
}}.
\]
The vector field $D_i^{(n)}$ satisfies a conservation law which can be obtained by differentiating Eq.~(\ref{eq:determinant_vector}) with respect to time \cite{anghelutaAnisotropicVelocityStatistics2012,mazenkoVelocityDistributionStrings1999}. 
This gives $ \partial_t D_i^{(n)}=-\epsilon_{ikl} \partial_k J_{l}^{(n)}$, with the associated current given by $J_{l}^{(n)} = \Im (\d_t \eta_n \d_l \eta_n^*)$. 
Thus
\[
(1) = -\epsilon_{ikl} \frac{6 \pi}{N q_0^2} \sum_{n=1}^N q_{j}^{(n)} (\partial_k J_l^{(n)}) \delta^{(2)}(\eta_n).
\]
Differentiating through the delta-function in the second term (2), we get 
\[D_i^{(n)} \partial_t \delta^{(2)} (\eta_n) =  \epsilon_{ikl} (\partial_k \eta_{n,1} ) (\partial_l \eta_{n,2})\sum_{r=1}^2 \left (\frac{\partial} {\partial  \eta_{n,r}} \delta^{(2)}(\eta_n)\right ) \partial_t \eta_{n,r}, \]
where $\eta_{n,1}$ and $\eta_{n,2}$ denotes the real and imaginary part of $\eta_n$, respectively.  
Straight forward, but tedious algebra, shows that this is equal to
\[
-\epsilon_{ikl} J_l^{(n)} \partial_k \delta^{(2)}(\eta_n) = -\epsilon_{ikl} \Im (\d_t \eta_n \d_l \eta_n^*) \sum_{r=1}^2 \left (\frac{\partial} {\partial  \eta_{n,r}} \delta^{(2)}(\eta_n)\right ) \partial_k \eta_{n,r},\]
after inserting $\eta_n = \eta_{n,1} + \I \eta_{n,2}$.
Thus
\[
(2) = -\epsilon_{ikl} \frac{6 \pi}{N q_0^2} \sum_{n=1}^N q_{j}^{(n)}  J^{(n)}_l \partial_k \delta^{(2)}(\eta_n).
\]
Taken together, this gives 
\[
\partial_t \alpha_{ij} = -\epsilon_{ikl} \partial_k
\left ( \frac{6 \pi}{N q_0^2} \sum_{n=1}^N q_{j}^{(n)}  J^{(n)}_l  \delta^{(2)}(\eta_n)
\right ),
\]
as desired.

\section{Calculation details of dislocation velocity}
\label{appendix:details_of_velocity_calculation}

Inserting the expression for the delta-function in terms of the dislocation density tensor $\delta^{(2)}(\eta_n) = \alpha_{ik} D_i^{(n)}q_k^{(n)}/ (2\pi |\vec D^{(n)}|^2)$ into Eq. \eqref{eq:dislocation_current_amplitudes}, we get
\[
\J_{lj} = \frac{6 \pi}{N q_0^2} \sum_{n=1}^N \alpha_{ik} J_{l}^{(n)} q_{j}^{(n)} \frac{D_i^{(n)} q_k^{(n)}}{2\pi |\vec D^{(n)}|^2}.
\]
Equating $\J_{lj}^{(\alpha)}$ and $\J_{lj}$ at a point $\vec r'$ on the dislocation line, where $\alpha_{ij} = t_i' b_j\delta^{(2)}(\Delta\vec r_\perp)$ using $\vec b \cdot \vec q^{(n)} = 2\pi s_n $,
\[
\epsilon_{lmn} t_m' b_j v_n' \delta^{(2)}(\Delta\vec r_\perp)
=
\frac{6\pi}{N q_0^2} \sum_{n=1}^N s_n t_i'  \delta^{(2)}(\Delta\vec r_\perp) J_{l}^{(n)} q_{j}^{(n)} \frac{D_i^{(n)} }{ |\vec D^{(n)}|^2}.
\]
We now integrate out the delta-function in the $\mathcal N'$-plane and contract both sides of the equation with $b_j$ to get
\[
\epsilon_{lmn} t_m' v_n' 
=
\frac{12 \pi^2}{N q_0^2|\vec b|^2} \sum_{n=1}^N s_n^2   \frac{t_i' D_i^{(n)} }{|\vec D^{(n)}|^2} J_{l}^{(n)},
\]
as desired.

\section{Amplitude decoupling}
\label{appendix:amplitude_decoupling}

The (complex) polynomial $f^{\rm s}$ (see Eq.~\eqref{eq:energyamplitude}) results from the amplitude expansion of the $\psi^3$ and $\psi^4$ terms in Eq.~\eqref{eq:F_PFC}. It may be computed by substituting Eq.~\eqref{eq:psi_approx} into Eq.~\eqref{eq:F_PFC} and integrating over the unit cell, under the assumption of constant amplitudes
\cite{goldenfeldRenormalizationGroupApproach2005,athreyaRenormalizationgroupTheoryPhasefield2006,salvalaglioCoarseGrainedModeling2022}. It features terms reading $\prod_{\ell=1}^{L} \eta_{n_\ell}$, with $L=3,4$ and $n_\ell$ for which the condition $\sum_{\ell=1}^L$ $\mathbf{q}^{(n_\ell)}=0$ is satisfied. By multiplying this condition by $\mathbf{b}$ and using Eq.~\eqref{eq:s_n} it then follows that 
\[\label{eq:conditiondecoupling}
\sum_{\ell=1}^L s_{n_\ell}=0.
\]
In the equation for the dislocation velocity, Eq.~(\ref{eq:3Ddislocationvelocity}), the only contributing amplitudes are those for which $s_n\neq 0$. The condition \eqref{eq:conditiondecoupling} implies that at least one of the other amplitudes, $\{\eta_m\}_{m\neq n}$, appearing in terms of $f^s$ containing $\eta_n$, also has $s_m\neq 0$ and then vanishes at the corresponding defect. Thus, for a given amplitude $\eta_{n}$ with $s_{n} \neq 0$, the terms in $\frac{\partial f^s}{\partial \eta_n^*}$ always contain at least one vanishing amplitude. Eq.~\eqref{eq:amptimefuncder} then reduces to Eq.~\eqref{eq:amplitude_evolution_equation_simple} at the defect as $\eta_n=0$ and $\frac{\partial f^s}{\partial \eta_n^*}=0$ there. Importantly, a full decoupling of the evolution equation for amplitudes which vanish at the defect is obtained.

This can be straightforwardly verified for specific lattice symmetries and dislocations. When accounting for the bcc lattice symmetry through $\mathbf{q}^{(n)}$ as in Eq.~\eqref{eq:qn_bcc}, the (complex) polynomial $f^{\rm s}$ entering the coarse-grained energy $F_\eta$ defined in Eq.~(\ref{eq:energyamplitude}) is
\begin{equation}
    f^{\rm s}=
    -2T(
    \eta_1^* \eta_2  \eta_6
    +\eta_1^* \eta_3  \eta_5
    +\eta_2^*  \eta_3 \eta_4
    +\eta_4  \eta_5^* \eta_6)
    +6V(\eta_1^* \eta_2 \eta_4^*  \eta_5
    +\eta_1^* \eta_3  \eta_4  \eta_6
    +\eta_2^* \eta_3  \eta_5  \eta_6^*)+\text{c.c.}
\end{equation}
which gives 
\begin{equation}
\begin{split}
    \frac{\partial f^{\rm s}}{\partial \eta_1^*}=-2T(\eta_2\eta_6+\eta_3\eta_5)+6v(\eta_2 \eta_4^*  \eta_5+\eta_3 \eta_4 \eta_6),\\
    \frac{\partial f^{\rm s}}{\partial \eta_2^*}=-2T(\eta_1\eta_6^*+\eta_3\eta_4)+6V(\eta_1 \eta_4  \eta_5^*+\eta_3 \eta_5 \eta_6^*),\\
    \frac{\partial f^{\rm s}}{\partial \eta_3^*}=-2T(\eta_1\eta_5^*+\eta_2\eta_4^*)+6V(\eta_1 \eta_4^*  \eta_6^*+\eta_2 \eta_5^*  \eta_6),\\
    \frac{\partial f^{\rm s}}{\partial \eta_4^*}=-2T(\eta_2\eta_3^*+\eta_5\eta_6^*)+6V(\eta_1^* \eta_2  \eta_5+\eta_1 \eta_3^*  \eta_6^*),\\
    \frac{\partial f^{\rm s}}{\partial \eta_5^*}=-2T(\eta_1\eta_3^*+\eta_4\eta_6)+6V(\eta_1 \eta_2^*  \eta_4+\eta_2 \eta_3^* \eta_6),\\
    \frac{\partial f^{\rm s}}{\partial \eta_6^*}=-2T(\eta_1\eta_2^*+\eta_4^*\eta_5)+6V(\eta_1 \eta_3^*  \eta_4^*+\eta_2^* \eta_3 \eta_5).
\end{split}
\label{eq:dfda}
\end{equation}
By comparing Eqs.~(\ref{eq:dfda}) with the dislocation charges for the possible Burgers vector in the bcc lattice, Table \ref{tab:dislocharges}, and noting that, \textit{at} the dislocation core, $\eta_n=0$ for $s_n\neq 0$, we find
\[
s_n\neq 0: \quad \frac{\partial f^{\rm s}}{\partial \eta_n} = 0,
\]
allowing for a decoupled system of evolution relations for $\eta_1,\cdots, \eta_6$, as described by Eq.~\eqref{eq:amplitude_evolution_equation_simple}.

\bibliography{Bibliography}

\end{document}